\documentclass[prd,aps,showpacs,twocolumn,nofootinbib]{revtex4-1}

\usepackage{amsmath,amssymb,amsthm,wasysym,amsfonts}
\usepackage{graphicx}
\usepackage{psfrag}
\usepackage{epstopdf}
\usepackage{color}
\usepackage{mathrsfs}
\usepackage{fixmath}

\newcommand{\bse}{\begin{subequations}}

\def\lm{{\ell m}}

\def\baralpha{\bar{\alpha}}

\def\de{\partial}
\def\f{\hat{f}}
\def\lm{{\ell m}}

\def\ii{{\rm i}}

\def\fc{{f^{\rm c}}}

\def\f{{\hat{f}}}
\def\C{{\cal C}}

\def\M{{\cal M}}

\def\k{\kappa}

\def\k{{\kappa_2^T}}

\newcommand{\be}{\begin{equation}}
\newcommand{\ee}{\end{equation}}

\usepackage{float}
\definecolor{cyan}{rgb}{0,0.9,0.9}
\definecolor{orange}{rgb}{0.9,0.5,0}
\definecolor{magenta}{rgb}{1,0,1}
\definecolor{purple}{rgb}{0.8,0.4,0.8}

\begin{document}
\title{Measurability of the tidal polarizability of neutron stars in late-inspiral gravitational-wave signals}
\author{Thibault \surname{Damour}}
\author{Alessandro \surname{Nagar}}
\affiliation{Institut des Hautes Etudes Scientifiques, 91440
  Bures-sur-Yvette, France}
\affiliation{ICRANet, 65122 Pescara, Italy}

\author{Lo\"\i c \surname{Villain}}
\affiliation{Laboratoire de Math\'ematiques et de Physique Th\'eorique,\\ Univ. F. Rabelais - CNRS (UMR 7350), F\'ed. Denis Poisson, 37200 Tours, France}

\begin{abstract}
The gravitational wave signal from a binary neutron star inspiral contains information
on the nuclear equation of state. This information is contained in a combination of the 
tidal polarizability parameters of the two neutron stars and is clearest in the late
inspiral, just before merger. We use the recently defined tidal extension of the effective
one-body formalism to construct a controlled analytical description of the frequency-domain
phasing of neutron star inspirals {\it up to merger}.
Exploiting this analytical description we find that the tidal polarizability parameters of
neutron stars {\it can be measured} by the advanced LIGO-Virgo detector network from 
gravitational wave signals having a reasonable signal-to-noise ratio of $\rho=16$. This
measurability result seems to hold for all the nuclear equations of state leading to a 
maximum mass larger than $1.97M_\odot$. We also propose a promising new way of extracting 
information on the nuclear equation of state from a coherent analysis of an ensemble of
gravitational wave observations of separate binary merger events.
\end{abstract}
\date{\today}


\maketitle

\section{Introduction}
Binary neutron star (BNS) inspirals are among the most promising sources for 
the advanced version of the ground based gravitational wave (GW) detector
network LIGO-Virgo.
BNS's evolve under the influence of gravitational radiation reaction
leading to a GW inspiral signal whose amplitude increases up to the merger,
while its frequency also increases up to a merger frequency $f_{\rm merger}\lesssim 2000$~Hz.
One of the goals of the observation of GW signals from BNS systems is to 
improve our knowledge about neutron star (NS) structure and the highly uncertain 
equation of state (EOS) of NS matter.
Advanced LIGO is expected to be able to detect about 40 BNS merger events per
year~\cite{Abadie:2010cf} with signal to noise ratio (SNR) $\rho\geq 8$.
The question that we shall address here is whether such observations can allow us to
learn something useful about the EOS of neutron star matter via the measurement
of {\it tidal polarizability} parameters from  the inspiral signal.

In Newtonian gravity the (quadrupolar) tidal polarizability of a body is usually measured by means
of the dimensionless Love number $k_2$ such that $\mu_2=2/(3G)k_2 R^5$, where $R$ denotes the
radius of the NS,  yields the ratio between the tidally induced quadrupole moment $Q_{ab}$
and the companion's perturbing tidal gradient $G_{ab}=\de_{ab}U$.
The generalization of the concept of tidal Love number $k_2$ to strongly self-gravitating
objects (NS or black holes) was discussed long ago by one of us as part of
the theory of motion of compact bodies~\cite{Damour:1983}.
This work indicated how, by matching a quadrupolar deformed NS geometry treated \`a la 
Thorne and Campolattaro~\cite{thorne:1967}, one could compute $k_2$ for a given neutron star EOS.
Recently, an explicit, simple, way of doing this matching computation of $k_2$ has
been obtained by Hinderer~\cite{Hinderer:2007mb}. The resulting numerical values for $k_2$ obtained
in Ref.~\cite{Hinderer:2007mb} have then been used in a preliminary analysis of the measurability
of tidal effects in BNS GW inspiral signals~\cite{Flanagan:2007ix}.
However, this early work has been marred by a calculational error in~\cite{Hinderer:2007mb} leading
to a substantial overestimate of the value of $k_2$. 
Later work~\cite{Damour:2009vw,Binnington:2009bb} emphasized that $k_2$ is a strongly 
decreasing function of the NS compactness $\C\equiv GM/(c^2R)$, such that $k_2(\C)$ formally vanishes 
in the black hole limit\footnote{It was already 
mentioned in Ref.~\cite{Damour:1983} that the $k_2$ of a (four-dimensional) black hole vanishes.
See~\cite{Kol:2011vg} for a discussion of black-hole Love numbers in higher spacetime dimensions.} 
$\C\to 1/2$, and generalized the computation of Love numbers so as to include gravito-magnetic
tidal polarizability coefficients as well as higher multipolar contributions.
Recently~\cite{Hinderer:2009ca} the tidal polarization parameter\footnote{We follow the notation for tidal
polarizability parameters introduced some time ago in the 
General Relativistic Celestial Mechanics formalism~\cite{PhysRevD.45.1017}: 
namely $\mu_\ell$ for the $\ell^{\rm th}$-multipolar {\it mass-type} (gravito-electric) coefficient, and $\sigma_\ell$ 
for the corresponding {\it spin-type} (gravito-magnetic) one.} $\mu_2$ was computed 
for a wide range of EOS. Moreover,  the question of discriminating between NS EOSs via 
GW observations with the advanced LIGO-Virgo detector network, using the early part 
(frequencies $f<450$~Hz) of the inspiral signal,  has been discussed 
and answered in the negative in~\cite{Hinderer:2009ca}: only if one has a 
GW signal with very high SNR $\rho=35$, and if the actual EOS of NS matter is 
unusually stiff, can one start distinguishing (at the $68\%$ confidence level) 
the early-inspiral tidal signal from the noise.

The reason why Refs.~\cite{Flanagan:2007ix,Hinderer:2009ca} performed a conservative
data analysis based only on the early inspiral GW signal, $f<450$~Hz, was that 
their analysis was based on using a purely post-Newtonian (PN) expanded description 
of the phasing, without having any way of controlling the validity of this description
for frequencies above 450~Hz. More precisely, they use a  
TaylorF2-type~\cite{Damour:2000zb} description of the frequency-domain GW phase 
of the form $\Psi(f) = \Psi^0_{\rm PN}(f)+\Psi^T_{\rm N}(f)$, with a point-mass
phasing  $\Psi^0_{\rm PN}(f)$ treated at 3.5PN accuracy~\cite{Damour:2000zb} 
and with a tidal phasing $\Psi^T_{\rm N}(f)$ treated at leading, 
Newtonian order~\cite{Flanagan:2007ix}.

Recently, a new, improved description of the dynamics, waveform and phasing  of compact 
binary systems has been developed based on the effective one body (EOB) 
formalism~\cite{Buonanno:1998gg,Buonanno:2000ef,Damour:2000we,Damour:2001tu,Damour:2008gu,Damour:2009ic}.
In particular, the way to extend the EOB formalism so as to include tidal effects has been
presented in~\cite{Damour:2009wj}. Let us recall that the EOB formalism is an analytical framework 
which combines several different theoretical results and approaches, and, in particular, contains {\it resummed} versions of the 
usual PN-expanded results. Such a framework has proven to be a powerful tool for constructing 
analytic waveforms that agree with numerical simulations. In the binary black hole case, 
EOB waveforms are in agreement with high-accuracy numerical waveforms at the remarkable 
level of 0.01~rad up to merger~\cite{Damour:2009kr,Pan:2009wj,Pan:2011gk}. 
In addition, the tidal-EOB formalism of~\cite{Damour:2009wj} has been successfully compared 
to state-of-the-art numerical simulations of BNS systems~\cite{Baiotti:2010xh,Baiotti:2011am}.
This comparison showed that the tidal-EOB formalism could reproduce the numerical phasing essentially
{\it up to merger} within numerical uncertainties.

This successful comparison (together with recent analytical progress~\cite{Bini:2012gu} in the computation
of the EOB tidal interaction potential)
 motivates us here to use the tidal-EOB formalism as a way to define 
a controlled analytical description of the phasing of tidally interacting BNS systems up to merger. 
More precisely, we will show below that the tidal contribution $\Psi^T_{\rm EOB}(f)$ 
to the Fourier domain phase predicted by the tidal-EOB approach can be represented 
(within less than 0.3~rad) by a certain (PN-type) analytical expression up to merger. This will allow
us to perform a data analysis using the full tidal phasing signal up to merger, while
keeping the convenience of having an explicit analytical representation of the 
tidal phasing (instead of the well-defined, but more indirect, full EOB description of tidally
interacting BNS systems).
Using such a EOB-controlled description of the tidal phasing up to merger, we will show 
(see Fig.~4 below) that the EOS-dependent tidal polarizability parameters $G\mu_2$ of NSs 
{\it can be measured}, at the $95\%$ confidence level, with the advanced LIGO-Virgo detector 
network using GW signals with reasonable SNRs ($\rho=16$) for all EOS in the sample we shall 
consider (only restricted by the observational constraint of yielding
a maximum mass larger than $1.97M_\odot$~\cite{Demorest:2010bx}).
In addition we shall propose a new way of extracting EOS-dependent information from a coherent
analysis of a collection of GW observations of separate BNS merger events, which promises a
large increase in measurement accuracy.

In this paper we will focus on BNS systems, but the formalim we present can be used as it is
for discussing the measurability of tidal parameters in mixed BH-NS binary systems.
This would allow one to go beyond the recent works~\cite{Pannarale:2011pk,Lackey:2011vz} 
dealing with some aspects of  the measurability of tidal polarizability coefficients from 
mixed binary systems.

The paper is organized as follows: in Sec.~\ref{sec:anlyt} we will review the main elements 
of the tidal-EOB formalism and present the analytic tidal phasing model in the frequency domain
that we will use in estimating the measurability of $\mu_2$. The theoretical aspects of our
measurability analysis are given in Sec.~\ref{sec:measure_theory}. 
The numerical results for the measurability of $\mu_2$ are presented 
in Sec.~\ref{sec:measure_results}, while concluding remarks are gathered 
in Sec.~\ref{sec:end}. 
The paper is completed by two Appendices. In Appendix~\ref{sec:eob} we extend and complete
the review of the tidal-EOB formalism of Sec.~\ref{sec:anlyt}, giving in particular the 
explicit analytical expressions for the tidal corrections to the EOB waveform. 
Finally, Appendix~\ref{sec:PN_general} collects the PN-expanded formulas for the tidal phasing for
a general relativistic binary that are used in the main text.
When convenient, we use geometrized units with $G=c=1$.

\section{Analytical tidal phasing models in the frequency domain}
\label{sec:anlyt}

The main aim of the present paper will be to estimate the measurability of
tidal parameters by making use of the full BNS inspiral signal, including the 
late-inspiral part just before merger, where tidal effects are strongest. 
We will do so by taking advantage of the recent development of an analytical
model which can accurately describe the full inspiral signal up to merger.
Indeed, in Refs.~\cite{Baiotti:2010xh,Baiotti:2011am} state-of-the-art 
numerical simulations of inspiralling BNS systems were compared to several 
analytical models. It was found that the EOB model (in its tidally extended
version as defined in~\cite{Damour:2009wj}) was able to match the numerical 
results up to merger. 
The EOB model (dynamics and waveform) is originally defined in the {\it time domain}.
For the data-analysis purpose of the present paper it will be convenient to
have in hands an {\it analytic} representation of the waveform in the 
{\it frequency domain}. The derivation of such an analytic frequency-domain
phasing model will be the topic of the present Section.

\subsection{Tidal effects in EOB dynamics}

Let us recall that the EOB 
formalism~\cite{Buonanno:1998gg,Buonanno:2000ef,Damour:2001tu} 
consists of three main elements: (i) a resummed Hamiltonian 
describing the conservative dynamics; (ii) a radiation-reaction 
force computed from the instantaneous angular momentum loss; 
(iii) a resummed waveform. 

For a nonspinning binary black hole (BBH) system of masses $M_A$, $M_B$ 
the EOB Hamiltonian is given by
\begin{equation}
\label{eq:Heob}
H_{\rm EOB}(r,p_{r_*},p_\varphi) \equiv M c^2\sqrt{1+2\nu (\hat{H}_{\rm eff}-1)} \  ,
\end{equation}
where
\begin{equation}
\label{eq:Heff}
\hat{H}_{\rm eff} \equiv \sqrt{p_{r_*}^2 + A(r) \left( 1 +
  \frac{p_{\varphi}^2}{r^2} 
+ z_3 \, \frac{p_{r_*}^4}{r^2} \right)} \, .
\end{equation}
Here $M\equiv M_A + M_B$ is the total mass, $\nu \equiv M_A \, M_B /
(M_A + M_B)^2$ is the symmetric mass ratio, and $z_3 \equiv 2\nu
(4-3\nu)$. In addition, we are using rescaled dimensionless
(effective) variables, namely $r \equiv r_{AB}c^2 / GM$ and
$p_{\varphi} \equiv P_{\varphi}c / (GM_A M_B)$, and $p_{r_*}$ is
canonically conjugated to a ``tortoise'' modification of
$r$~\cite{Damour:2009ic}.
The crucial input entering this Hamiltonian is the ``radial potential''
$A(r)$, whose leading-order approximation is $A(r)\approx 1-2/r+\dots\equiv 1-2GM/(c^2 r_{AB})+\dots$

The proposal of Ref.~\cite{Damour:2009wj} for including dynamical
tidal effects in the conservative part of the dynamics consists in
using a tidally-augmented radial potential of the form
\begin{equation}
\label{eq:A}
A(r) = A^0(r) + A^{\rm tidal} (r)  .
\end{equation}
where $A^0(r)$ is the point-mass potential defined in
Eq.~\eqref{eq:3.4} of Appendix~\ref{sec:eob}, while $A^{\rm tidal}(r)$ 
is a supplementary ``tidal contribution'' describing the tidal
interaction potential.
In terms of the dimensionless gravitational potential 
$u\equiv GM/(c^2 r_{AB})\equiv 1/r$ it reads
\begin{align}
\label{eq:At}
A^{\rm tidal}=\sum_{\ell\geq 2} - \kappa_\ell^T u^{2\ell+2}\hat{A}^{\rm tidal}_\ell(u).
\end{align}
Here the term $\kappa^T_{\ell} u^{2\ell +2}$ represents the multipolar
tidal interaction of degree $\ell$, taken at Newtonian order in a 
PN expansion.
The dimensionless EOB tidal parameter $\kappa_\ell^T$ entering Eq.~\eqref{eq:At}
is related to the {\it tidal polarizability} coefficients $G\mu_\ell^{A,B}$ 
of each neutron star as~\cite{Damour:2009wj}
\begin{align}
{\kappa}_\ell^T &\equiv \kappa^A_\ell + \kappa_\ell^B 
\end{align}
where
\be
\label{eq:def_kappaA}
\kappa_\ell^A\equiv \left(2\ell - 1\right)!!\dfrac{X_B}{X_A}\dfrac{G\mu_\ell^A}{(GM/c^2)^{2\ell+1}}.
\ee
where we recall that $M=M_A+M_B$ denotes the total mass of the binary
and $X_A\equiv M_A/M$.
The tidal polarizability coefficient $G\mu_\ell^A$ has the 
dimension $[\rm length]^{2\ell+1}$. It measures the ratio 
between the $\ell$-th multipole moment induced in body $A$ 
and the external tidal gradient felt by body $A$. Among these
multipolar tidal polarizability coefficients, the dominant one 
is the quadrupolar, $\ell=2$, one, $G\mu_2^A$. [Note that $\mu_2$ is
denoted by $\lambda$ in Refs.~\cite{Flanagan:2007ix,Hinderer:2009ca,Pannarale:2011pk}].
In addition, if $R_A$ denotes the radius of body $A$, $G\mu_\ell^A$ is related to the corresponding dimensionless 
Love number $k_\ell^A$ by
\be
(2\ell-1)!!\,G\mu^A_\ell = 2 k_\ell^A \, R_A^{2\ell+1},
\ee
so that
\be
\kappa_\ell^A = 2 k_\ell^A \dfrac{X_B}{X_A}\left(\dfrac{R_A}{G(M_A+M_B)/c^2}\right)^{2\ell+1}.
\ee
The additional factor $\hat{A}^{\rm tidal}_\ell(u)$ in Eq.~\eqref{eq:At} 
represents the effect of distance-dependent, higher-order relativistic 
contributions to the dynamical tidal interactions: 1PN, i.e. first order 
in $u$, 2PN, i.e. of order $u^2$, etc.
Here we will use the following ``Taylor-expanded'' form of $\hat{A}^{\rm tidal}_\ell$
\begin{equation}
\label{eq:Ahat}
\hat{A}^{\rm tidal}_\ell(u)=1+\bar{\alpha}_1^{(\ell)} u + \bar{\alpha}_2^{(\ell)} u^2 \ ,
\end{equation}
where $\bar{\alpha}_n^{(\ell)}$ are functions of $M_A$, ${\cal C}_A$,
and $k_\ell^A$ for a general binary and are  defined as  
(see Eq.~(37) of~\cite{Damour:2009wj})
\be
\label{eq:bar_alpha1}
\bar{\alpha}_n^{(\ell)}\equiv \dfrac{\kappa_\ell^A\alpha^{A(\ell)}_n + \kappa_\ell^B\alpha^{B(\ell)}_n}{\kappa_\ell^A + \kappa_\ell^B},
\ee
where $\alpha^{A(\ell)}_n$ is the coefficient of the the $n$PN fractional correction
to the tidal interaction potential of body $A$.
(see Sec.~IIIC of~\cite{Damour:2009wj}). 
The individual dimensionless coefficient $\alpha_n^{A(\ell)}$ is a function of the
dimensionless ratio $X_A\equiv M_A/M$. [Note that $X_B\equiv M_B/M=1-X_A$].
The analytical expression of the first post-Newtonian, quadrupolar 
($\ell=2$) coefficient $\alpha_1^{A(2)}$ has been reported in~\cite{Damour:2009wj} 
(and then confirmed in~\cite{Vines:2010ca}) and reads
\be
\label{eq:alpha1}
\alpha_1^{A(2)}=\dfrac{5}{2}X_A.
\ee
Recently, Ref.~\cite{Bini:2012gu} has succeeded in computing the first post-Newtonian
octupolar ($\ell=3$) coefficient  $\alpha_1^{A(3)}$, as well as the {\it second post-Newtonian} 
quadrupolar ($\ell=2$) and octupolar ($\ell=3$) coefficients $\alpha_2^{A(\ell)}$.
The most relevant 2PN quadrupolar coefficient reads
\be
\label{eq:alpha2}
\alpha_2^{A(2)}= \dfrac{337}{28}X_A^2 + \dfrac{1}{8} X_A + 3.
\ee
In the equal-mass case, $X_A=1/2$, the values of these coefficients are 
$\alpha_1^{A(2)}=\bar{\alpha}^{(2)}_1=5/4=1.25$ and $\alpha_2^{A(2)}=\bar{\alpha}_2^{(2)}=85/14\approx 6.071429$. 
A recent comparison~\cite{Baiotti:2010xh,Baiotti:2011am} between 
EOB predictions and BNS numerical simulations concluded 
that $\bar{\alpha}^{(2)}_2\lesssim 40$.
In the following, we shall restrict ourselves to considering {\it only} tidal
quadrupolar contributions, i.e. we will take only the $\ell=2$ value in
Eqs.~\eqref{eq:At} and~\eqref{eq:Ahat}. 
It is shown in Section~\ref{sec:effect_of_ell} of Appendix~\ref{sec:eob}
that the effect of higher-$\ell$ tidal corrections is small. It will be 
neglected in our analysis.

\subsection{EOB waveform and its stationary phase approximation}
\label{EOB:SPA}

When considering tidally interacting binary systems, one needs to augment the
point-mass waveform  $h_{\lm}^0$ by tidal contributions. Similarly to the additive 
tidal modification~\eqref{eq:At} of the $A$ potential, we will here consider 
an {\it additive} modification of the waveform, having the structure
\begin{equation}
\label{eq:4.1}
h_{\lm} =  h_{\lm}^0  + h_{\ell m}^{\rm tidal}.
\end{equation}
See Appendix~\ref{sec:eob} for the explicit expressions of $h_{\lm}^0$ 
and  $h_{\ell m}^{\rm tidal}$.
In turn, this tidally modified waveform defines a corresponding tidally modified
radiation reaction force ${\cal F}_\varphi $ through its instantaneous angular 
momentum loss.

The radiation-driven EOB dynamics defined by $H_{\rm EOB}(A)$ and ${\cal F}_\varphi$
(where both $A$ and ${\cal F}_\varphi$ are tidally modified) allows us to compute
a  time-domain multipolar GW signal $h_\lm(t)=A_{\lm}(t)e^{-\ii \phi_\lm(t)}$.
Following Refs.~\cite{Baiotti:2010xh,Baiotti:2011am}, we characterize
the (time-domain) phasing of the quadrupolar waveform $h_{22}(t)$ by means of the 
following function of the instantaneous quadrupolar GW frequency 
$\omega=\omega(t)\equiv d\phi/dt$ [where $\phi(t)\equiv \phi_{22}(t)$]
\be
\label{eq:def_Qomg}
Q_\omega(\omega) \equiv \dfrac{d\phi(t)}{d\ln\omega(t)}\equiv \dfrac{[\omega(t)]^2}{\dot{\omega}(t)}.
\ee
In the stationary phase approximation (SPA), the phase $\Psi(f)$ of the 
frequency-domain waveform $\tilde{h}(f)$,  i.e. the phase of the Fourier 
transform of the time-domain (quadrupolar) waveform,
\be
\label{eq:h22_t}
\tilde{h}_{22}(f) \equiv \tilde{A}(f) e^{-\ii \Psi(f)},
\ee
is simply the Legendre transform of the quadrupolar time-domain phase $\phi(t)$, namely
\be
\label{eq:psi_f}
\Psi_{\rm SPA}(f) = 2\pi f t_f - \phi(t_f)-\dfrac{\pi}{4},
\ee
where $t_f$ is the saddle point of the Fourier transform, i.e. the solution of the
equation $\omega(t_f)=2\pi f$.
Differentiating Eq.~\eqref{eq:psi_f} twice with respect to $f$ leads to the
following link between $\Psi_{\rm SPA}(f)$ and the function $Q_\omega(\omega)$
\be
\label{eq:SPA_tot}
\dfrac{d^2\Psi_{\rm SPA}(\omega_f)}{d\omega_f^2} = \dfrac{Q_\omega(\omega_f)}{\omega_f^2},
\ee
where $\omega_f$ now denotes the Fourier domain circular frequency $\omega_f\equiv 2\pi f$.
Below we will simply denote the Fourier domain frequency $\omega_f$ as $\omega$
without bothering to distinguish it from the time-domain $\omega(t)$.

In the following we shall decompose the result~\eqref{eq:SPA_tot} 
in its point-mass and tidal parts, thereby relating the ``tidal part'' 
of the Fourier-domain phase $\Psi_{\rm SPA}(f)$ to
the ``tidal part'' of $Q_\omega(\omega)$. 
On the one hand, the tidal part, say $Q_\omega^{T}(\omega)$,  of $Q_\omega(\omega)$ 
is computed as
\be
Q_\omega^T(\omega) = Q_\omega(\omega)-Q_\omega^0(\omega),
\ee
where  $Q_\omega^0(\omega)$ is the outcome of a point-mass EOB simulation, i.e., 
one without tidal effects in both the dynamics and the waveform.
Then, the corresponding tidal part,
\be
\label{eq:PsiT_EOB}
\Psi^T_{\rm EOB}(\omega) =  \Psi_{\rm EOB}(\omega) - \Psi^0_{\rm EOB}(\omega),
\ee
of the Fourier-domain EOB phase satisfies, within the SPA approximation, the relation
\be
\label{eq:EOBSPA}
\dfrac{d^2\Psi^T_{\rm EOB_{SPA}}(\omega)}{d\omega^2}=  \dfrac{Q^T_\omega(\omega)}{\omega^2}.
\ee
Let us emphasize that we expect the SPA approximation to the phasing to remain
accurate {\it up to the merger}. Indeed the small parameter that controls the 
validity of the SPA is essentially $\epsilon_{\rm adiab}=\dot{\omega}/\omega^2\equiv 1/Q_\omega$.
For instance, Ref.~\cite{Damour:2000gg}, Eqs.~(3.9)-(3.10), has computed 
the next contribution beyond  the leading SPA and found that it introduces 
a dephasing $\delta\Psi$ which, 
in the case of Newtonian chirps, is equal to $\delta \Psi=(23/24)\times (2/9)\times 1/Q_\omega$.
The quantity $Q_\omega$ is very large during early inspiral and decreases towards
the merger. Looking at the value of the full $Q_\omega = Q_\omega^0 + Q_\omega^T$ 
in the exact EOB description of tidally interacting BNS systems, we have checked that the 
(equal-mass) value of $Q_\omega(\omega)$ for $\omega=\omega^{\rm contact}$
(where $\omega^{\rm contact}$ is the EOB approximation to the merger frequency, see below)
remains larger than
about 20 for realistic compactnesses ($\C=0.14-0.18$). Though this value is 
reduced (by $\sim 10$) from the corresponding point-mass value  $Q_\omega^0(\omega_{\rm contact})$,
it is still comfortably large compared to $1$, so that one can expect the phasing error
linked to the use of the SPA to be a small fraction of a radian.

\subsection{PN-expansion of the EOB tidal phasing}
\label{sec:PNaccuracy}
In Sec.~\ref{sec:measure_results} below we shall estimate the measurability 
of the tidal parameter
$\kappa_2^T$ by computing the Fisher matrix ${\bf F}$ corresponding to the simultaneous
measurement of a tidal parameter, say $\lambda_T\propto \kappa_2^T$, with several
other, non tidal, parameters, say $\lambda_a$, $a=1,\dots,n$. Though, in principle
we could numerically compute the relevant Fisher matrix ${\bf F}$ by evaluating 
the numerical derivatives of the full, Fourier domain, EOB waveform $\tilde{h}_{\rm EOB}(f;\,\lambda_T,\lambda_a)$
with respect to all the parameters $(\lambda_T,\lambda_a)$, it will be convenient
to estimate ${\bf F}$ by replacing the numerically computed $\tilde{h}_{\rm EOB}(f;\,\lambda_T,\lambda_a)$
(which involves computing the numerical Fourier transform of a numerically generated
time-domain EOB waveform $h_{\rm EOB}(t;\,\lambda_T,\lambda_a)$) by some sufficiently
accurate analytic approximation. We will do so by combining several approximations,
the validity of which we shall control.
The first approximation we shall use is the SPA, which we have discussed in the
previous section. 
The second approximation will consist in using post-Newtonian expansions
to derive adequately accurate expressions of the two parts of the Fourier
domain phase
\be
\label{eq:Psi_FT}
\Psi(f) = \Psi^0(f;\,\lambda_a) + \Psi^{T}(f;\,\lambda_T,\lambda_a).
\ee
In this section we study how many terms in the PN expansion of 
the tidal phase $\Psi^{T}(f;\,\lambda_T,\lambda_a)$ we must retain
to obtain an approximation to $\Psi^T(f)$ which remains reasonably
close to the EOB prediction {\it up to merger}.
From Eq.~\eqref{eq:PsiT_EOB} we see (in the SPA) that to answer this
question we need to compare the PN-expansion of the tidal part of $Q_\omega^T(\omega)$
of $Q_\omega(\omega)$ to the ``exact'' value of $Q_\omega^T(\omega)$ defined by
the EOB model. 
During most of the inspiral , not only is the phase evolution quasi-adiabatic,
i.e. $Q_\omega \gg 1$, as already discussed above, but the dynamical evolution
can also be well approximated by an adiabatic quasi-circular inspiral.
In the latter approximation, the function $Q_\omega=\omega^2/\dot{\omega}$
is obtained by writing the balance equation between the instantaneous energy flux $F$ at
infinity and the adiabatic evolution of the energy of the system (i.e., the Hamiltonian, $H(\omega)$)
expressed as a function of the instantaneous GW frequency $\omega=2\Omega$ (where $\Omega$
is the orbital frequency). This yields $-F = dH/dt=\left(dH/d\omega\right) \dot{\omega}$, from which
one obtains
\be
\label{eq:QomgHF}
Q_\omega^{\rm adiab}(\omega) = - \omega^2 \dfrac{dH(\omega)/d\omega}{F(\omega)}.
\ee 
Reexpressing this result in terms of the dimensionless rescaled angular momentum
$j\equiv J/(G\mu M)$, the Newton normalized energy flux $
\hat{F}\equiv F/F^{\rm Newton}\equiv {\cal F}_\varphi/{\cal F}^{\rm Newton}_\varphi$, and
replacing the independent variable $\omega$ by the usual, dimensionless PN ordering
parameter
\be
x  \equiv \left(\dfrac{1}{2} \dfrac{GM\omega}{c^3}\right)^{2/3}\equiv \left(\dfrac{\pi GM f}{c^3}\right)^{2/3}
\ee
leads to an expression of the form
\be
\label{eq:Qomg_total}Q^{\rm adiab}_\omega(x;\,\nu,\lambda_T) = \dfrac{5}{48\nu}x^{-5/2}b(x;\,\nu,\lambda_T)
\ee
where the function $b(x;\,\nu,\lambda_T)$, defined as
\be
b(x;\,\nu,\lambda_T) = -2 x^{3/2}\dfrac{\de_xj(x;\,\nu,\lambda_T)}{\hat{F}(x;\,\nu,\lambda_T)},
\ee
is simply equal to 1 in the Newtonian approximation. More precisely, it 
starts as
\be
b(x;\,\nu,\lambda_T) = 1 + \dfrac{1}{336}\left(743 + 924\nu\right)x - 4\pi x^{3/2} + {\cal O}(x^2).
\ee
Starting from the adiabatic EOB dynamics, the function $j(x)$ is obtained by 
eliminating $u=1/r$ between the EOB expression $j^2(u) = -A'(u)/(u^2 A(u))'$
(obtained by minimizing the effective potential for circular orbits $A(r)(1+j^2/r^2)$)
and the expression of $\Omega$ in terms of $u$ obtained from the Hamilton equation
$\Omega = \de H/\de J$.[See Sec.~III and IV of Ref.~\cite{Damour:2009wj} and 
Appendix~\ref{sec:eob} for more details about the EOB circular dynamics].
On the other hand, the function $\hat{F}(x)$ is obtained by
as a sum of various resummed circular multipolar waveforms $h_\lm(x)$ 
of Ref.~\cite{Damour:2008gu}.

In the following, we shall replace the (EOB-resummed) adiabatic approximation 
$Q_\omega^{\rm adiabatic}$,  Eq.~\eqref{eq:QomgHF}, to $Q_\omega^{\rm EOB}$, by a
sufficiently accurate PN expansion of $Q_\omega^{\rm adiabatic}$. This is to avoid
an inaccurate feature of $Q_\omega^{\rm adiabatic}$ during the late inspiral. From Eq.~\eqref{eq:QomgHF}
we see that $Q_\omega^{\rm adiabatic}(\omega)$ is proportional to the derivative
$dH(\omega)/d\omega$ which, by construction, vanishes at the Last Stable Orbit (LSO),
where the circular energy $H(\omega)$ reaches a minimum.
By contrast, the exact $Q_\omega^{\rm EOB}$ does not vanish at the LSO, nor the PN
expanded version of $Q_\omega^{\rm EOB}$ that we shall use. The frequency 
corresponding to the tidal-EOB defined LSO happens to be quite close to the
contact frequency. Using  $Q_\omega^{\rm adiabatic}$ up to contact might then introduce
inaccuracies in the phasing just before merger. Our use below of a suitable 
PN-expanded representation of $Q_\omega$ avoids this source of uncertainty and
maintains consistency with the SPA by allowing the value of $Q_\omega$ at contact
to remain of order 20 for all cases considered.

Current analytical knowledge that has been incorporated in the EOB description 
of tidal effects~\cite{Baiotti:2010xh,Baiotti:2011am} allows us to compute the 
tidal part $Q_\omega^T(\omega)$ of $Q_\omega(\omega)$ and therefore, using 
Eq.~\eqref{eq:EOBSPA}, the tidal part $\Psi^T(\omega)$ of the Fourier domain
phase $\Psi(\omega)$ beyond the 1PN accuracy obtained in Ref.~\cite{Vines:2011ud}.  
First, the fact that the EOB formalism  naturally accomodates the inclusion of 
tail effects in the waveform allows us to obtain a PN-expanded tidal 
phasing model that is analytically complete up to 1.5PN order. [In addition, the
EOB formalism already contains the next order tail effects at 2.5PN order].
Second, the EOB approach is designed in a way which makes it easy to complete it
beyond current analytical knowledge by using effective field theory methods.
In particular, Ref.~\cite{Bini:2012gu} recently computed the 2PN tidal contributions to
the EOB radial potential $A(u)$, i.e. the coefficient $\bar{\alpha}_2$ of $u^2$ 
in Eq.~\eqref{eq:Ahat} (see Eq.~\eqref{eq:alpha2}). 
As mentioned in Ref.~\cite{Bini:2012gu}, a straightforward extension of the method
used to derive the 2PN tidal contribution to $A(u)$ can allow one to derive the 2PN
tidal contribution to the waveform. However this calculation has not yet been completed.
Waiting for this result, we shall here use the natural flexibility of the EOB formalism,
to parametrize the 2PN tidal corrections to the multipolar waveform by means of some
parameters that we will call $\beta_n^\lm$.
Let us recall that in order to obtain $Q_\omega^T$ at, say, the fractional 2.5PN accuracy,
the energy flux $F$  must be computed by retaining all the  $\ell=2$ and $\ell=3$ 
multipolar contributions to the waveform.  Then, to obtain the 
flux to 2.5PN accuracy we need the quadrupolar $\ell=m=2$ waveform 
(stripped of its tail factor) to 2PN fractional accuracy, and the odd-parity 
$\ell=2$, $m=1$ and $\ell=m=3$ and $\ell=3$, $m=1$ even-parity waveforms at
1PN fractional accuracy. Following Refs.~\cite{Damour:2009wj,Baiotti:2011am}, 
we shall parametrize such higher-PN tidal corrections to the waveform 
along the following model
\be
h_\lm^{\rm tidal} = h_\lm^{A\,{\rm tidal}} +h_\lm^{B\,{\rm tidal}},
\ee
with
\be
h_\lm^{A\,{\rm tidal}} = h_\lm^{A\, {\rm tidal_{Newt}}}\left(1 + \beta^{\lm}_1(X_A) x + \beta^{\lm}_2(X_A) x^2+\dots\right).
\ee
For the time being, the only such PN fractional tidal  waveform correction
which is known is $\beta_1^{22}(X_A)$. Using the 1PN-accurate results of 
Ref.~\cite{Vines:2011ud} one can indeed derive the following explicit
analytical expression
\be
\beta_1^{22}(X_A) = \dfrac{-202 + 560 X_A - 340 X_A^2 + 45 X_A^3}{42(3 - 2 X_A)}.
\ee
However, at 2.5PN, the final result depends on several
other higher-corrections, namely $\beta_2^{22}(X_A)$, $\beta_1^{21}(X_A)$,
$\beta_1^{33}(X_A)$ and $\beta_1^{31}(X_A)$ , that account respectively 
for 2PN fractional tidal corrections to the $\ell=m=2$ multipole and for 
1PN fractional tidal corrections to the $\ell=2$, $m=1$ and $\ell=3$, $m=1$ 
and $m=3$ subdominant multipoles.
In Appendix~\ref{sec:PN_general} we present the explicit expressions 
for $Q_\omega^T$
and $\Psi^T$ at 2.5PN accuracy for the general case of unequal mass binary
systems. In the text below we shall specify those general formulas to the 
particular,
but physically most relevant, case of equal-mass neutron star binaries
(having therefore equal-compactnesses and equal tidal parameters).

In the equal-mass case, because of symmetry reasons, the only higher-order tidal 
waveform parameter that contributes to the phasing is $\beta_2^{22}$.
We then arrive at the following explicit expression for the 2.5PN 
accurate tidal contribution to $Q_\omega$
\begin{align}
\label{eq:QomgT_equal}
Q_\omega^T(x) &= -\k \dfrac{65}{6} x^{5/2}\bigg[1 
+ \dfrac{4361}{624}x - 4\pi x^{3/2}\nonumber\\
& + \left(\dfrac{4614761}{122304}+\dfrac{4}{3}\bar{\alpha}_2^{(2)}+\dfrac{4}{13}\beta_2^{22}\right)x^2 
-\dfrac{4283}{156}\pi x^{5/2}\bigg],
\end{align}
where we recall that $\bar{\alpha}^{(2)}_{22}=\alpha_{2}^{A(2)}(X_A=1/2)=85/14$ 
and where $\beta_2^{22}$ denotes the value of the (unknown) function 
$\beta^2_{22}(X_A)$ for $X_A=1/2$.
Using Eq.~\eqref{eq:EOBSPA}, the corresponding 2.5PN accurate tidal phase of the 
Fourier transform  of the GW signal reads (in the SPA)
\begin{align}
\label{eq:2.5PN}
\Psi^T_{\rm 2.5PN}(x)&=-\k \dfrac{39}{4} x^{5/2}
\bigg\{1+\dfrac{3115}{1248}x - \pi x^{3/2} \nonumber\\
& + \left(\dfrac{23073805}{3302208}+\dfrac{20}{81}\baralpha_2^{(2)}+\dfrac{20}{351}\beta_2^{22}\right)x^2\nonumber\\
& - \dfrac{4283}{1092}\pi x^{5/2}\bigg\}.
\end{align}
Such an explicit representation of a Fourier domain phase as a polynomial in 
$x(f)\propto f^{2/3}$ is usually called TaylorF2~\cite{Damour:2000zb}.
The 2.5PN TaylorF2 formula~\eqref{eq:2.5PN} improves the 1PN result of 
Ref.~\cite{Vines:2011ud} in that: (i) tail effects are 
included up to 2.5PN order; (ii) a large part of the 2PN term is explicitly computed, 
although it still depends on the yet uncalculated quantity  $\beta_2^{22}$ 
(2PN tidal correction to the waveform).
Note that at leading, Newtonian, order, Eq.~\eqref{eq:2.5PN} predicts that the (equal-mass) 
tidal dephasing at contact, i.e. for $x_{\rm contact}=\C_A=\C_B$ (see Eq.~\eqref{eq:x_contact})
is of order 
\be
\label{eq:psi_cont}
\Psi_{\rm Newt-equal-mass}^{\rm contact}=-\dfrac{39}{32}\dfrac{k_2^A}{\C_A^{5/2}},
\ee
which, for the typical values $k_2=0.08$ and $\C=1/6$, yields $-8.6$~rad. With the further 
amplification of PN effects discussed below
this means that the tidal dephasing at contact is of order $-10$~rad 
(see Fig.~\ref{fig:EOB_vs_PN_phase}).

Let us now indicate why we expect that the contribution to the tidal phase coming from 
$\beta_2^{22}$ is likely to be numerically subdominant compared to the currently known
terms. Let us first note that at leading, Newtonian order the overall coefficient 
$(39/4) \kappa_2^T$ in the tidal phase Eq.~\eqref{eq:2.5PN}, is, in view of 
Eq.~\eqref{eq:QomgHF}, the sum of a tidal contribution from the Hamiltonian 
$H$ and a tidal contribution from the energy flux $F$. More precisely, one finds
that 
\be
\dfrac{39}{4} = \dfrac{3}{4}\left(9_H + 4_F\right)
\ee
where the indices indicate the origin ($H$ or $F$) of the contribution. Already at this
leading-order level, one notices that the contribution from the energy flux is subdominant
(by a factor 2.25) with respect to the contribution from the Hamiltonian, i.e. from the
radial potential $A(u)$.
When pursuing this analysis at the 1PN level and considering the {\it fractional} PN 
modification of the tidal phase $\hat{\Psi}^T_{\rm 2.5PN}\equiv \Psi^T_{\rm 2.5PN}/ \Psi^T_{\rm Newt} \equiv 1 + {\cal O}(x) $ 
in Eq.~\eqref{eq:2.5PN} one finds (still for the equal-mass case)
\begin{align}
\hat{\Psi}^T_{\rm 1PN}&= 1 + \left(\dfrac{125795}{61152} + \dfrac{55}{156}\bar{\alpha}_1^{(2)} + \dfrac{10}{91}\beta_1^{22}\right)x\nonumber\\
                    &\approx 1 + \left(2.05709 + 0.35256 \bar{\alpha}_1^{(2)} + 0.10989\beta_1^{22}\right) x,
\end{align}
where we decomposed the 1PN fractional contribution into three parts: i) one coming from the leading
order tidal terms in $H$ and $F$;  ii) one coming from the 1PN tidal correction to $H$ (term $\propto \bar{\alpha}_1^{(2)}$);
and (iii) one coming from the 1PN tidal correction to the quadrupolar waveform (and flux, term $\propto \beta_1^{22}$).
In the equal-mass case one has $\bar{\alpha}_1^{(2)}=5/4=1.25$ and $\beta_1^{22}=-11/672\approx -0.0164$.
As a consequence of these numerical values, we see that: a) the coefficient of $\beta_1^{22}$ is 
smaller by a factor 3.2 than the coefficient of $\bar{\alpha}_1^{(2)}$; 
b) in addition, as the numerical value of $\beta_1^{22}$ is $\approx -0.0164$, its contribution to
the total 1PN fractional coefficient is  $0.10989\times 0.0164/2.05709 \approx 8.8 \times 10^{-4}$ 
times smaller than the first term 2.057 and $7.2\times 10^{-4}$ times smaller than the sum of the first
two contributions.
Performing a similar analysis at the 2.5PN level (now inserting the known numerical values 
of $\bar{\alpha}_1^{(2)}$ and $\beta_1^{22}$) yields  
\begin{align}\label{eq:2.5PNhat}
\hat{\Psi}^T_{\rm 2.5PN}&\approx  1 + 2.50 x -\pi x^{3/2} \\
                      &+ (6.99 + 0.25 \bar{\alpha}_2^{(2)} + 0.057 \beta_{2}^{22})x^2 - 3.92 \pi x^{5/2}\nonumber.
\end{align}
Again we see that the contribution from $\beta_2^{22}$ is likely to be subdominant. 
Indeed, not only is the coefficient of $\beta_2^{22}$  4.3 times smaller than the one
of $\bar{\alpha}_2^{(2)}$, but it is also about 149 times smaller than the known 
2PN coefficient $6.99+0.25\bar{\alpha}_2^{(2)}\approx 8.51$.
Independently of these numerical arguments, let us note that, as already mentioned above, 
an important feature of
the adiabatic approximation to the function $Q_\omega(\omega)$ is that it vanishes at 
the adiabatic LSO. 
Since tidal effects strongly influence the LSO location (see Ref.~\cite{Damour:2009wj}) this indicates
that tidal corrections to the Hamiltonian (i.e., to $A(u)$) have a dominant influence on
the shape of the $Q_\omega(\omega)$ function below the LSO frequency and thereby on the 
tidal corrections to $\Psi^T(\omega)$.
In view of these arguments, in the following we will neglect the effect of $\beta_2^{22}$ 
both in the exact EOB phase and its 2.5PN approximant, $\Psi^T_{\rm 2.5PN}(\omega)$.
We will then work with Eq.~\eqref{eq:2.5PN} with $\beta^{22}_2=0$ (but $\bar{\alpha}_2^{(2)}=85/14\approx 6$) 
as a numerically acceptable approximation to the 2.5PN tidal phase.

\subsection{Accuracy of PN-expanded representations of the EOB phasing}
\label{test:PsiT_EOB}
Let us now study to what extent the ``exact'' EOB tidal phasing $\Psi_{\rm EOB_{SPA}}^T$ 
obtained by integrating Eq.~\eqref{eq:EOBSPA} with the exact (time-domain)
$Q_\omega(\omega)$ defined by Eq.~\eqref{eq:def_Qomg} on the right hand side,
can be represented by various PN expansions. 
\begin{figure}[t]
\center
\includegraphics[width=0.5\textwidth]{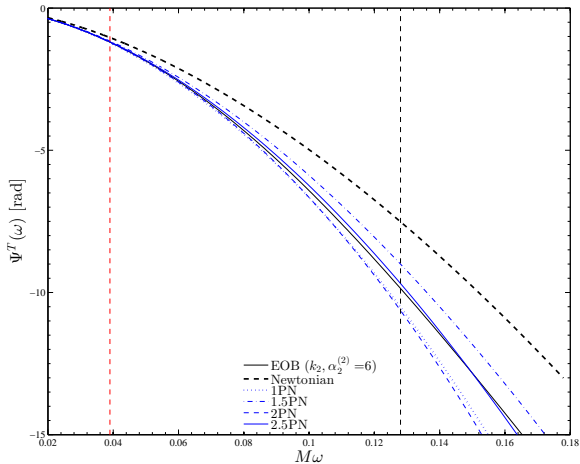}
\caption{\label{fig:EOB_vs_PN_phase} Successive PN approximants to the (SPA) tidal contribution 
to the EOB phase as function of the $\ell=m=2$ GW frequency. The leftmost vertical line indicates
450~Hz for a $1.4M_\odot + 1.4M_\odot$ BNS system. The rightmost vertical line indicates the contact
frequency that is taken as a fiducial analytical definition of the moment of merger.
The plot refers to a $\C=0.16$, $\gamma=2$ polytropic model.}
\end{figure}
Fig.~\ref{fig:EOB_vs_PN_phase} illustrates the performance of Eq.~\eqref{eq:2.5PN} 
(considered at different orders of truncation) in reproducing $\Psi_{\rm EOB_{SPA}}^T$. 
The figure refers to an equal-mass binary, with $\gamma=2$ polytropic EOS $p=K\mu^\gamma$,
with compactness ${\cal C}=0.16$. In addition, each star has $M_A=1.47838M_\odot$, $R_A=13.6438$~km,
$k_2^A=0.063122$ so that the EOB dimensionless tidal parameter of the system is equal 
to $\kappa_2^T=75.2476$, which yields $G\mu_2 = 19896\,{\rm km}^5$.
As mentioned above, we use $\beta_2^{22}=0$, and $\bar{\alpha}_2^{(2)}=6$ for simplicity.
The phase $\Psi^T_{\rm {EOB}_{SPA}}(\omega)$ is computed by integrating numerically 
Eq.~\eqref{eq:EOBSPA}
starting from the frequency $\omega_0$  that marks the beginning of the 
inspiral waveform obtained when solving the EOB equations of motion numerically. 
This integration is done using the 2.5PN result for $\Psi^T$  and $d\Psi^T/d\omega$ 
as initial boundary conditions, and thus $\omega_0$ needs to be chosen  sufficiently 
small (i.e., the EOB inspiral waveform has to be sufficiently long) so 
to have $Q_\omega^{\rm 2.5PN}\approx Q_\omega^{\rm EOB}$. We refer the reader to 
Appendix~\ref{sec:eob_theory} to get further technical  details related to 
Fig.~\ref{fig:EOB_vs_PN_phase}. 

The various PN approximations gathered in Fig.~\ref{fig:EOB_vs_PN_phase} are obtained from
Eq.~\eqref{eq:2.5PN} and are represented as: thick dashed line (black online) for the Newtonian, 
dotted line (blue online) for the 1PN, dash-dotted 
line (red online) for the 1.5PN, dashed line (red online) for the 2PN and 
solid line (red online) for the full, 2.5PN phase. The leftmost vertical line
indicates the frequency 450~Hz 
(used as cutoff in Refs.~\cite{Flanagan:2007ix,Hinderer:2009ca}), while
the rightmost vertical line indicates the frequency of ``bare'' contact, that 
defines within the EOB formalism the merger frequency. This bare contact is 
defined as the GW frequency where the relative distance $R=M/u\approx M/x$
is equal to the sum of the radii of the two NS $R_A+R_B=M_A/{\cal C}_A+M_B/{\cal C_B}$,
i.e.
\be
\label{eq:x_contact}
\dfrac{1}{x_{\rm contact}}=\dfrac{X_A}{\C_A}+\dfrac{X_B}{\C_B}
\ee
from which the gravitational wave frequency at contact is computed using 
$M\omega_{\rm contact}=2\pi M f_{\rm contact}=2 (x_{\rm contact})^{3/2}$. In the equal-mass
case Eq.~\eqref{eq:x_contact}  yields the simple result $x_{\rm contact}=\C_A=\C_B$.

Among the useful informations contained in this figure let us note that:
i) the Newtonian approximation substantially differs from the EOB phase
even at low frequencies, and exhibits a discrepancy of about 3 rad at
merger; ii) as the PN order $n$ is increased, the convergence towards the EOB
prediction is non monotonic and the sign of the difference $\Delta_n=\Psi_{\rm nPN}-\Psi_{\rm EOB}$
alternates as $n$ takes the successive values $0,1,1.5,2$. This is linked to the alternating
signs in $\hat{\Psi}^T_{\rm 2.5PN}$ in Eq.~\eqref{eq:2.5PN}. In particular the difference 
$\Delta_1$ reaches the value $-0.6$ rad at contact; 
iii) it is only at  2.5PN accuracy that we get a rather accurate representation 
of the EOB tidal phase. 
Note that at merger, where the frequency parameter $x$ reaches the value $x^{\rm contact}=\C=0.16$, 
the fractional PN modification of the tidal phase is equal to 

\be
\hat{\Psi}^T_{\rm 2.5PN}(x^{\rm contact})\approx  1 + 0.40_1 - 0.20_{1.5} + 0.22_2 -0.13_{2.5}\approx 1.29
\ee
which illustrates the effect of the successive PN approximations, labelled here by the corresponding 
PN order, $(1,1.5,2,2.5)$.

\begin{figure}[t]
\center
\includegraphics[width=0.5\textwidth]{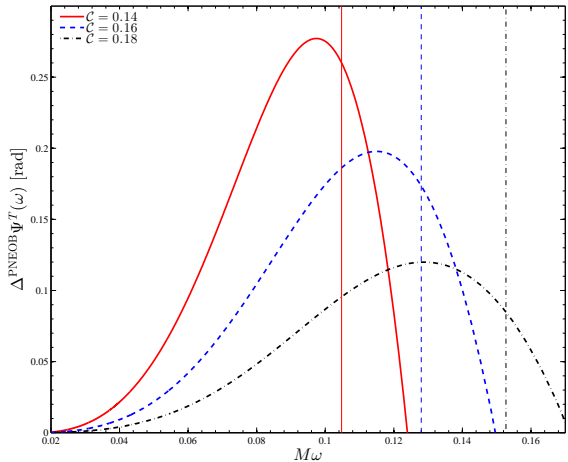}
\caption{\label{fig:fig2} Difference between the 2.5PN-expanded tidal (Fourier) phase and the
corresponding exact EOB one obtained by integrating Eq.~\eqref{eq:EOBSPA}. 
Each curve refers to a $\gamma=2$ polytropic model with different compactness. 
The vertical lines indicate the corresponding contact frequency.}
\end{figure}

To firm up the conclusions drawn from the particular model (with compactness ${\cal C}=0.16$) 
considered in Fig.~\ref{fig:EOB_vs_PN_phase}, we studied what happens when the compactness varies
within a realistic range.
Let us recall that the magnitude of the dimensionless EOB tidal parameter $\kappa_2^T=\kappa_2^A+\kappa_2^B$ 
is related (in the equal-mass case) to the Love number $k_2$ and to the compactness $\C$ by
\be
\kappa_2^T=\dfrac{1}{8}\dfrac{k_2}{\C^5}.
\ee
[For a general multipolar index one has $\kappa_\ell^T=1/(2^{2\ell-1})k_\ell/\C^{2\ell+1}$].
For a given EOS, as ${\cal C}$ increases, $k_2$ decreases in a correlated 
manner~\cite{Damour:2009vw,Hinderer:2009ca},
so that $\kappa_2^T$ varies by about a factor 9 in a range of realistic compactnesses.
For instance, in the case of the $\gamma=2$ polytrope that we are currently discussing, 
as $\C$ varies between 0.14 and 0.19, $\kappa_2^T$ decreases from 183.37 down to 21.757,
with radii correspondingly varying from 14.369~km down to 12.435~km.
We generalized the comparison reported in Fig.~\ref{fig:EOB_vs_PN_phase} for three
different compactnesses  ${\cal C}=\{0.14,0.16,0.18\}$.
The results for the differences 
\be
\Delta^{\rm PNEOB}\Psi^T(\omega) = \Psi^T_{\rm 2.5PN}(\omega)-\Psi^T_{\rm EOB}(\omega)
\ee
are shown in Fig.~\ref{fig:fig2}. In all cases these differences are rather small,
being $\lesssim 0.3$ rad at merger. In addition, let us note that the positive 
sign of $\Delta^{\rm PNEOB}\Psi^T$ (given the fact that $\Psi^T$ is negative) 
means that using $\Psi^T_{2.5PN}$ instead of the more exact EOB phasing $\Psi^T_{\rm EOB}$ 
is a {\it conservative} way of estimating the measurability of tidal parameters.

\section{Measurability of tidal parameters: theoretical discussion}
\label{sec:measure_theory}

\subsection{Fisher matrix formalism}

Under usual simplifying assumptions (Gaussian noise, sufficiently high SNR) 
the variance $\sigma_{\k}^2$ in the measurement of $\k$ is computed using the
standard Fisher matrix formalism, as already used in the context of binary
systems in Refs.~\cite{Cutler:1994ys,Poisson:1995ef,Flanagan:2007ix,Hinderer:2009ca,Pannarale:2011pk}.
When considering a Fourier--domain waveform $\tilde{h}(f;\,\lambda_i)$ that is a 
function of $n+1$ parameters $\{\lambda_i\}=\{\lambda_a,\lambda_T\}$,  
the Fisher information matrix is a $(n+1)\times(n+1)$ matrix whose elements are
given by 
\be
F_{ij}\,=\,\langle\de_{\lambda_i}h ,\de_{\lambda_j}h\rangle
\ee
where $\langle,\rangle$ denotes the Wiener scalar product between two signal 
$\tilde{h}(f)$ and $\tilde{k}(f)$, defined as
\be
\langle h,k\rangle \equiv 4\,\Re\,\int\limits_0^{+\infty} df \dfrac{\tilde{h}^*(f)\tilde{k}(f)}{S_n(f)},
\ee
with $S_n(f)$ denoting the one-sided strain noise of the detector. 
In absence of specific prior, the variance in the measurement of 
each parameter $\lambda_i$ is given by the corresponding  diagonal 
element of the  inverse Fisher matrix (or covariance matrix), 
\be
\sigma_{\lambda_i}^2=(F^{-1})^{ii}.
\ee
Assuming the SPA approximation and neglecting relativistic corrections to the
amplitude, the Fourier transform of the waveform is
\be
\label{eq:FT}
\tilde{h}(f) = A f^{-7/6}e^{-\ii \Psi(f)}\theta(f^{\rm contact}-f),
\ee
where $\theta(f^{\rm contact}-f)$ denotes a Heaviside step function indicating
that we cut off the inspiral signal above the contact frequency defined above in 
Eq.~\eqref{eq:x_contact}.[ This is a coarse approximation to the post-merger 
signal that might be improved by extending the EOB representation to an effective
description of the post-contact GW signal]. 
The amplitude parameter $A$ has been shown to be uncorrelated to 
the other parameters~\cite{Cutler:1994ys,Poisson:1995ef}, 
so that we shall forget about it in the following\footnote{However, strictly speaking, 
because of the dependence of the step-function $\theta(f^{\rm contact}-f)$  in
Eq.~\eqref{eq:FT} on the dynamical parameters via $f^{\rm contact}(\lambda_a)$,
there will be a small correlation between the amplitude parameter $A$ and the 
other parameters. Following~\cite{Cutler:1994ys}, we shall neglect this correlation 
which is not expected to modify our conclusions in any significant way.}.

From this equation, the squared signal to noise ratio (SNR) is written as
\be
\label{eq:SNR}
\rho^2 = 4\,\int\limits_0^{+\infty} df \dfrac{|\tilde{h}(f)|^2}{S_n(f)}= 4\,A^2\int\limits_0^{\fc} df \dfrac{f^{-7/3}}{S_n(f)},
\ee
where $f^c\equiv f^{\rm contact}$, and the elements of the Fisher matrix are 
\be
F_{ij}= 4\,A^2\,\int\limits_0^{\fc} \dfrac{df}{S_n(f)} f^{-7/3}\de_{\lambda_i}\Psi \de_{\lambda_j}\Psi
\ee
In view of the common proportionality of both $\rho^2$ and $F_{ij}$ to the (randomly vaying) 
squared signal amplitude $A^2$, it is convenient to define the following ``reduced'' 
Fisher matrix 
\be
\hat{F}_{ij}\equiv \dfrac{F_{ij}}{\rho^2}.
\ee
One then sees that this reduced Fisher matrix can be written as
\be
\label{red_fish}
\hat{F}_{ij}=\int\limits_0^\fc d f \,\gamma( f)\,\de_{\lambda_i}\Psi( f)\de_{\lambda_j}\Psi(f),
\ee
whre $\gamma(f)df$ denotes the following measure
\be
\gamma(f)df \equiv \dfrac{ df\,f^{-7/3} S_n^{-1}(f) }
                           { \int\limits_0^{\fc} df\,f^{-7/3} S_n^{-1}(f)}.
\ee
Note that this measure is normalized to unity, $\int\limits_0^{\fc}\gamma(f)d f=1$. 
This measure naturally leads to defining a new (Euclidean) scalar product 
among {\it real} phase functions
\be
\label{renorm_prod}
(a|b)\equiv \int\limits_0^\fc d f \gamma(f)\,a(f)\,b(f),
\ee
in terms of which we can write the rescaled Fisher matrix as
\be
\hat{F}_{ij}=\left(\de_{\lambda_i}\Psi|\de_{\lambda_j}\Psi\right).
\ee
The elements of the inverse of this reduced Fisher matrix then
give the ``SNR-normalized probable errors'', $\rho\,\sigma_{\lambda_i}$  
on each parameter, namely
\be
\label{sigma_hat}
\hat{\sigma}_{\lambda_i}\equiv\rho\,\sigma_{\lambda_i} = \sqrt{(\hat{F}^{-1})^{ii}}.
\ee
In the following $S_n$ is taken to be the ${\tt ZERO\_DET\_high\_P}$ anticipated sensitivity curve 
of Advanced Ligo~\cite{shoemaker}.
The minimum of the effective dimensionless strain noise $fS_n(f)$ for this sensitivity curve 
is located at frequency $f_0=56.56$~Hz. In the following we shall often work with the reduced
frequency parameter $\f \equiv f/f_0$. 

\subsection{Phasing model and parameter dependence}

Concretely, we shall use a Fourier domain waveform of the type of Eq.~\eqref{eq:FT},
with a phase $\Psi(f)$ in the form
\be
\Psi(f) = \Psi^0(f) + \Psi^T(f),
\ee
where, as above, $\Psi^0(f)$ denotes the point-mass contribution to the SPA phase and $\Psi^T(f)$
the tidal part. We shall approximate both contributions with some PN expansion.
As already discussed above, the tidal contribution will be approximated by the 2.5PN 
accurate expression of Eq.~\eqref{eq:2.5PN}. Concerning the point-mass phase, it is 
currently analytically known up to 3.5PN order~\cite{Damour:2002kr,Buonanno:2009zt}.
As we shall further discuss below, for the purpose of the present paper it will be
enough to use the following 2PN~\cite{Blanchet:1995ez} accurate representation of the point-mass phase
\begin{widetext}
\begin{align}
\label{eq:psi_0}
\Psi^0\left(\f;\,\lambda_1,\lambda_2,\lambda_3,\lambda_4,\beta,\sigma\right) & = \lambda_1 + 2 \pi \f \lambda_2 \nonumber\\
                                                                      &+ \frac{3}{128} (\pi \lambda_3 f_0 \f)^{-5/3} 
                                  \bigg[1 + \frac{20}{9} \left(\frac{743}{336} + \frac{11}{4} \lambda_4\right)  v_{\lambda_3,\lambda_4}^2\f^{2/3} 
                                -4 (4\pi-\beta) v_{\lambda_3,\lambda_4}^3\f\nonumber\\
                                & + 10\left(\dfrac{3058673}{1016064} + \dfrac{5429}{1008}\nu + \dfrac{617}{144}\nu^2-\sigma\right)v_{\lambda_3,\lambda_4}^4 \f^{4/3}\bigg],
\end{align}
\end{widetext}
where
\be
v_{\lambda_3,\lambda_4}\equiv\left(\pi\lambda_3\lambda_4^{-3/5}f_0\right)^{1/3}
\ee
and where the parameters  the $(\lambda_i)_{[1..4]}$ have the following meaning
\begin{align}
\label{eq:lambda1}
\lambda_1  &= -\phi_c - \pi/4,\\
\lambda_2 & = f_0 t_c, \\
\lambda_3 & = {\cal M}, \\
\label{eq:lambda4}
\lambda_4 & = \nu.
\end{align}
Here $\phi_c$ is a reference phase and $t_c$ a reference time, ${\cal M}\equiv\nu^{3/5} M$ is the chirp mass and 
$\nu=M_A M_B/M^2$ the symmetric mass ratio.
In addition, the parameter $\beta$ is a spin-orbit parameter and $\sigma$ a spin-spin one.

As for the (quadrupolar) tidal contribution $\Psi^T(f)$ we can write it in various forms depending on
the choice of tidal parameters we want to fit for. For instance,
if we choose as tidal parameter $\lambda_T$ determining the overall scale of the tidal phase 
the following symmetric combination of the two $\ell = 2$ tidal polarizability coefficients
\mbox{$G\mu_2^{ A,B} = (2/3) k^{A,B}_2 R_{A,B}^5$} (with the dimensions of $[\rm length]^5$),
\begin{align}
\label{eq:def-meanlamb}
\lambda_T & = G\bar{\mu}_2 
 \equiv \frac{G}{26} \left[\left(1 + 12 \frac{X_B}{X_A}\right) \mu_{2}^A + \left(1 + 12 \frac{X_A}{X_B}\right) \mu_{2}^B\right]\,,
\end{align}
we obtain a tidal signal of the form
\be
\label{eq:psi_T}
\Psi^T_{\rm 2.5PN}(\f;\,\lambda_T,\lambda_a) = - \lambda_T \frac{117}{8 M_{\lambda_3,\lambda_4}^5 \lambda_4}  
v_{\lambda_3,\lambda_4}^5 \f^{5/3} \hat{\Psi}^T_{\rm 2.5PN}(\f;\lambda_a),
\ee
where $M_{\lambda_3,\lambda_4}\equiv \lambda_3\lambda_4^{-3/5}$ denotes the total mass expressed as
a function of the chirp mass $\lambda_3$ and the symmetric mass ratio $\lambda_4$.
In this form the $\lambda_a$-dependence of the factor $v_{\lambda_3,\lambda_4}^5/(M_{\lambda_3,\lambda_4}^5 \lambda_4)$
introduces correlations when fitting for $\lambda_T$ together with the $\lambda_a$'s. 
An alternative choice might be to consider as tidal parameter the (dimensionless) combination 
\be
\label{eq:lambda_prime}
\lambda_T'= G\bar{\mu}_2 \frac{v_{\lambda_3,\lambda_4}^5}{M_{\lambda_3,\lambda_4}^5 \lambda_4},
\ee
so as to minimize the correlations when fitting $\lambda_T'$ together with the $\lambda_a$'s.
Note, however, that there will always remain correlations due to the $\lambda_a$-dependence
in the fractional PN correction factor $ \hat{\Psi}^T_{\rm 2.5PN}(\f;\lambda_a)$.
The $\lambda_a$-dependence of $\hat{\Psi}^T_{\rm 2.5PN}(\f;\lambda_a)$ has several sources: i) an
explicit dependence on $\M$ and $\nu$ coming through the argument $v=v_{\lambda_3,\lambda_4}\f^{1/3}$;
ii) an implicit dependence on the mass ratio coming from the individual Love numbers and the radii
entering the definition of $\hat{\Psi}^T_{\rm 2.5PN}$ (see Appendix~\ref{sec:PN_general}).
However, when computing the corresponding Fisher-matrix elements involving the partial derivative
with respect to $\lambda_a$ of $\Psi$, the contributions from the $\lambda_a$-dependence of the
tidal part $\Psi^T$ are largely subdominant compared to the large, early-inspiral dominating
contribution coming from $\de_{\lambda_a}\Psi^0$. 
We have indeed checked that taking into account the variability of the $\lambda_a$'s within
$\hat{\Psi}_{\rm 2.5PN}^T(\f;\,\lambda_a)$ or neglecting it only changes the error $\sigma_{\lambda_T}$
at the $1.5\times 10^{-3}$ fractional level. As for the $\lambda_a$ variability of the prefactor
of $\hat{\Psi}^T_{\rm 2.5PN}$ in Eq.~\eqref{eq:psi_T}, when using~\eqref{eq:def-meanlamb} as tidal
parameter, it was found (when $\beta$ and $\sigma$ are fixed, or well constrained), 
because of the signs of the correlations between $\lambda_T$ and $\lambda_a$,
to lead to a small, ${\cal O}(2\%)$, improvement in the measurability of $\lambda_T$ compared to that
of $\lambda'_T$ given by Eq.~\eqref{eq:lambda_prime}.
In the following we shall fit for $\lambda_T$, Eq.~\eqref{eq:def-meanlamb}, taking into account the
variability of the $\lambda_a$'s in the prefactor of the tidal phase [as displayed in Eq.~\eqref{eq:psi_T}],
however, for simplicity (and easier comparison with the unequal-mass case discussed below) we shall
neglect the variability of the $\lambda_a$'s entering the PN-correction factor $\hat{\Psi}^T_{\rm 2.5PN}(\f;\,\lambda_a)$,
(i.e. neglect their small contribution to $\de_{\lambda_a}\Psi^T$ computing the Fisher matrix).

In this work we keep in the tidal signal
$\Psi^T_{\rm 2.5PN}(f)$ only the contribution associated to the quadrupolar tidal deformation as measured
by $\kappa_2^A$ or $G\mu_2^A$. Actually, the EOB formalism takes into account higher multipolar
tidal interactions, as already done in previous work~\cite{Damour:2009wj,Baiotti:2010xh,Baiotti:2011am}.
Using this theoretical result, we show however in Appendix~\ref{sec:effect_of_ell} that the numerical contribution
of higher multipole moments ($\ell=3,4$) to the tidal signal is rather small ($\Delta\Psi^T\sim -0.2$ rad),
so that we are entitled to neglect it to estimate the measurability of $\k$. 
However, we recommend that in fitting real GW signals to tidal EOB templates one includes also the 
higher multipolar tidal contributions. But, in order not to introduce new parameters to be fitted, one should
express the higher-order polarizability parameters, $G\mu_\ell$, in terms of $G\mu_2$ only. 
More precisely, using $G\mu_2\sim k_2 R^5$, $G\mu_3\sim k_3 R^7$ and $G\mu_4\sim k_4 R^9$, one can 
reexpress $G\mu_3$ and $G\mu_4$ in terms of $G\mu_2$ and of the following 
combination of Love numbers
\begin{align}
k_{3/2}\equiv \dfrac{k_3}{k_2^{7/5}}\\
k_{4/2}\equiv \dfrac{k_4}{k_2^{9/5}}
\end{align}
e.g., $G\mu_3\sim k_{3/2}\, (G\mu_2)^{7/5}$. Replacing then the modified Love numbers $k_{3/2}$
and $k_{4/2}$ by some constant numbers (say of order 0.7 so as to approximately mimick 
the result of realistic EOSs) we end up with an approximate description of higher multipolar
contributions that is entirely expressed in terms of the quadrupole polarizability parameter
$G\mu_2$.

Let us now comment on the form, Eq.~\eqref{eq:psi_0}, of the point-mass phase that we shall use in this work.
This point-mass phase is only 2PN accurate. The reason for limiting our accuracy to this level is that, as 
we will see explicitly below, the terms in the Fisher matrix that determine the measurability of the  two dynamical
parameters entering the point-mass phase, namely the chirp mass $\lambda_3={\cal M}$ and the symmetric mass ratio
$\lambda_4=\nu$, are essentially\footnote{Here, for illustrative purposes, we keep only the leading-order PN signal 
contributing to the corresponding Fisher matrix element: e.g. $\de_{\lambda_3}\Psi \sim v^{-5}$ leading to
$\hat{F}_{33}\sim \left(v^{-5}|v^{-5}\right)=I_{-10}$.} proportional to integrals of the following 
types: $I_{-10}=\int d\ln f\,f \gamma(f) v(f)^{-10}$ and 
$I_{-6}=\int d\ln f\, f\gamma(f) v(f)^{-6}$.
While the integral giving the signal to noise ratio, 
Eq.~\eqref{eq:SNR} is proportional to $I_0=\int d\ln f f\gamma(f)$ 
and is roughly concentrated 
around a couple of frequency octaves around $\f=f/f_0=1$, 
the integrals $I_{-10}$ and $I_{-6}$ are mainly concentrated 
towards (different) lower frequencies.
\begin{figure}[t]
\center
\includegraphics[width=0.5\textwidth]{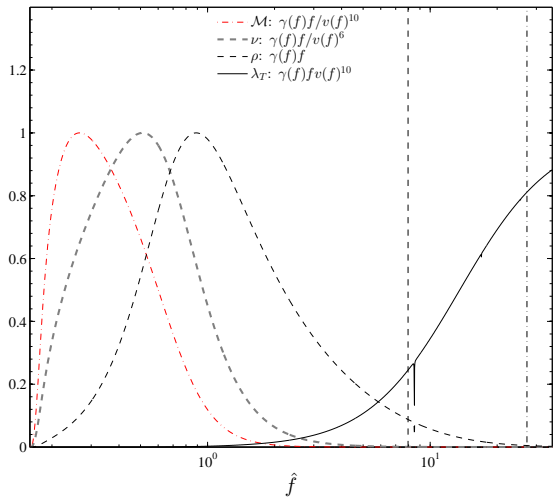}
\caption{\label{fig:gamma} Integrands, per frequency octave, of the integrals determining the measurability of 
${\M}$, $\nu$, $\rho$ (SNR) and $\lambda_T$. While most of the SNR is gathered around frequencies
\hbox{$\f=f/(56.56~{\rm Hz})\sim 1$}, the measurability of $\M$ and $\nu$ is concentrated
towards lower frequencies ($\f=f/f_0<1$), and that of the tidal parameter $\lambda_T$ gets its
largest contribution from the late inspiral up to the merger. The rightmost vertical line
indicates the merger frequency for $\C=0.1645$, while the leftmost vertical line marks 450~Hz
for a $1.4M_\odot + 1.4M_\odot$ BNS system.}
\end{figure}
The concentration on the logarithmic frequency axis of several relevant measurability signals
is  illustrated in Fig.~\ref{fig:gamma}. Note in particular how the integrands of $I_{-10}$ (chirp mass) 
and $I_{-6}$ (symmetric mass ratio) are peaked at frequencies below the SNR integrand of $I_0$.
Physically, this corresponds to saying that most of the useful cycles for the measurability of
${\cal M}$ and $\nu$ come from the early inspiral. As the PN expansion converges reasonably well
for such low frequencies, using a 2PN accurate phasing is guaranteed to be a reasonably good 
approximation for the point-mass part of the phase. This has been checked 
by Ref.~\cite{Poisson:1995ef} for the measurement of $\M$ and $\nu$,
which found (see their Table~II) that using a 2PN accurate (instead of
a 1.5PN accurate, as in Ref.~\cite{Cutler:1994ys}) template led to only  $\sim 10\%$ differences 
in the fractional uncertainties in $\nu$ and ${\cal M}$.
We found, as expected, that the situation is even better for the measurement of $\lambda_T$: namely,
we found that the fractional uncertainty on $\lambda_T$ is changed (and actually improved) when
using a 2PN template for $\Psi^0$, rather than a 1.5PN one, only at the $5\times 10^{-3}$ level
By contrast to the cases of $\M$ and $\nu$, the measurability of the tidal parameter $\lambda_T$ 
is associated in the Fisher matrix to an integral of the type 
$I_{+10}=\int d\ln f f\gamma(f) v(f)^{10}$, which gets its largest
contribution from the late inspiral up to the merger (see solid line in Fig.~\ref{fig:gamma}).
More specifically, the integrand of $I_{+10}$, i.e. $\propto f\gamma(f) f^{10/3}$ is equal to $f^2/S_n(f)$. 
The ${\tt ZERO\_DET\_high\_P}$ advanced LIGO noise curve $S_n(f)$ happens to be a rather flat
function of $f$ between $\sim 50$~Hz and $\sim800$~Hz and then increases to reach a shot noise 
behavior $S_n(f)\propto f^2$ at high frequencies. This implies that the integrand of 
$I_{+10}$,  i.e. $f^2/S_n(f)$, roughly grows like $f^2$ between 50~Hz and 800~Hz, to then asymptote
towards a finite limit at high frequencies.
The clear separation between, on the one hand, the two SNR curves associated to $\M$ and $\nu$
(which are relatively close to each other) and on the other hand the SNR curve associated to
$\lambda_T$ also indicates (as we shall discuss below) that $\M$ and $\nu$ are strongly
correlated among themselves, while $\lambda_T$ is not so strongly correlated to $\M$ and $\nu$.
The figure also displays two possible cut-off frequencies for the measurements of the tidal
signal: the conservative value 450~Hz (dashed vertical line) used in Refs.~\cite{Flanagan:2007ix,Hinderer:2009ca}, $\f=7.956$, 
or the compactness--dependent contact frequency that we shall use here, $(\pi M f )_{\rm contact}={\cal C}^{3/2}$
(dash-dotted vertical line, computed using EOS BSK21 with a model with $M=1.4M_\odot$ and ${\cal C}=0.1645$). 
Evidently, the use of the late-inspiral cut-off frequency  $\f_{\rm contact}$ calls for
a formalism able to describe the phasing up to the merger (here, the EOB formalism and its
accurate high PN expanded representation discussed in the previous section). 

In Eq.~\eqref{eq:psi_0} we have included also a parameter $\beta$ associated to the spin-orbit 
interaction and a parameter $\sigma$ associated to the spin-spin one~\cite{Blanchet:1995ez}. 
These parameters are equal to 
\begin{align}
\beta  &= \dfrac{1}{12}\left(113 X_A^2 + 75\nu\right){\bf\hat{L}} \cdot {\bf\hat{a}}_A + (A\leftrightarrow B),\\
\sigma &= \dfrac{\nu}{48}\left(-247{\bf\hat{a}}_A\cdot {\bf\hat{a}}_B + 721\, {\bf\hat{L}}\cdot {\bf\hat{a}}_A\, {\bf\hat{L}}\cdot {\bf\hat{a}}_B\right),
\end{align}
where ${\bf \hat{a}}_A={\bf S}_A/(G M_A^2)$ is the dimensionless spin parameter of body $A$.
Previous work~\cite{Cutler:1994ys,Poisson:1995ef,Hinderer:2009ca} discussing 
data-analysis including the spin parameters $\beta$ and $\sigma$  had incorporated 
Bayesian priors \`a la~\cite{Cutler:1994ys} 
constraining the magnitudes of $|\beta|$ and $|\sigma|$ to be smaller 
than $8.5$ and $5.0$ respectively, which are plausible theoretical upper limits 
on them. However, such values are very conservative bounds on $\beta$ and $\sigma$ 
in view of observed binary pulsar 
systems (as already pointed out in Refs.~\cite{Cutler:1994ys,Blanchet:1995ez}).
Indeed, recent estimates of the event-rate for BNS GW observations are mainly
obtained from extrapolation of the currently observed binary pulsar systems.
All the known binary pulsar systems have rather small observed spin parameters.
Considering the fastest spinning pulsar observed in a BNS system, namely PSR J0737-3039A,
whose spin period is 23~ms~\cite{Lorimer:2008se}, we concluded from the calculations of moments 
of inertia by Bejger~et~al.~\cite{Bejger:2005jy} (who work with the EOSs: BPAL12, APR, SLy, BGN2H1 and GNH3) 
and by Morrison~et~al.\cite{Morrison:2004df} (who use FPS), that the initial dimensionless 
spin parameter $\hat{a}$ is between approximatively $0.017$ (for BPAL12) and $0.03$ (for GNH3). 
This leads to an initial range for the corresponding parameter $\beta$ of order $|\beta|\,\in\,\left[0.11 ; 0.196\right]$, 
while the 2PN-level spin-spin parameter $\sigma$ is at most of the order $|\sigma|\lesssim 10^{-4}$. 
Taking into account the slowing down of the spin until the moment of merger, we estimated that $\beta$ 
at the time of the merger would be within the range $[0.09 ; 0.17]$ so that we decided to use the 
conservative upper limit of $0.2$ for $\beta$. Hence, we studied the measurability of $\beta$ 
together with the five other parameters $\{\lambda_{1,\dots,4};\,\lambda_T\}$ submitted to a 
Gaussian Bayesian prior $\propto \exp[-1/2(\beta/0.1)^2]$ constraining $|\beta|$ to be 
smaller than 0.2 at the $95\%$ confidence level. 
The result of the error estimates coming from such a constrained, six-parameter
Fisher matrix formalism will be presented in Table~\ref{table_beta} below, 
where they are compared to the result of a five-parameter Fisher matrix formalism where 
$\beta$ is set to zero from the beginning.
One sees from the numbers in Table~\ref{table_beta} that such a constrained six-parameter analysis
leads to only a very slight increase of the error estimates.
In view of this, in the following we shall neglect (i.e. set to zero from the start) $\beta$.
Similarly, and a fortiori, in view of the very small upper bound quoted above on $\sigma$, 
we can also neglect the 2PN level spin-spin parameter $\sigma$.
Let us emphasize that if, by contrast, one keeps the parameter $\beta$
while using the very conservative prior $\beta\leq 8.5$, this leads to a very large
increase of the error bars on $\M$ and $\nu$, and a noticeable increase of
the error bar on $\lambda_T$. 
As it will be exemplified in Table~\ref{table_beta}, the use
of the very conservative prior constraining $\beta\leq 8.5$ instead of the 
``realistic'' one leading to $\beta\leq 0.2$, increases the statistical measurement error 
on $G\mu_2$ by a factor which varies between 1.28 (for EOS BSK19) and 1.10 (for EOS GNH3).
In addition, if one does not neglect the spin-spin parameter $\sigma$ 
(as we shall do here), or alternatively, does not put a realistic prior on it, 
but instead fits for it using a seven parameter
Fisher matrix, constrained by the very conservative bound $|\sigma|<5.0$, 
this leads to a further, substantial increase of the measurement errors.

\section{Measurability of tidal parameters: numerical results}
\label{sec:measure_results}

\subsection{A sample of equations of state}

\begin{table*}[t]
  \caption{\label{tab:tableNS} Properties of a $M_A=1.4M_\odot$ neutron star for the 11 EOSs considered
in this paper. From left to right, the columns report: the name of the EOS;
the radius of the star $R$; its compactness ${\cal C}$; the value of the $\ell=2$ Love number $k_2$; 
the value of the tidal parameters $\k$ and $G\mu_2$; the value of the bare contact frequency 
$f^{\rm c}_{\rm bare}$ in Hz and the corresponding dimensionless contact frequency 
$\f^{\rm c}_{\rm bare}=f^{\rm c}_{\rm bare}/f_0$ with $f_0=56.56$~Hz.}
\begin{center}
\begin{ruledtabular}
\begin{tabular}{llllllll}
 EOS & $R$ [km] & ${\cal C}$  & $k_2$ & $\kappa_2^T$ & $G\mu_2~[{\rm km}^5]$ &  $f^{\rm c}$ [Hz] & $\f^{\rm c}$ \\
\hline
MS1   & 14.92   & 0.1390 &     0.1100 &   264.9895 &   53360.36  & 1196.09  & 21.15 \\
GNH3  & 14.19   & 0.1457 &     0.0852 &   162.099 &   32641.6  & 1284.06    & 22.702      \\  
MS2   & 13.71   & 0.1510 &     0.0883 &   140.6002 &   28312.35  & 1354.28  & 23.94 \\
BSK21 & 12.57   & 0.1645 &     0.0930 &    96.4647 &   19424.9  & 1540.29   & 27.23 \\  
MPA1  & 12.47   & 0.1660 &     0.0924 &    91.6308 &   18451.49  & 1561.00  & 27.60 \\
AP3   & 12.09   & 0.1710 &     0.0858 &    73.3528 &   14770.89  & 1632.06  & 28.85 \\
BSK20 & 11.75   & 0.1760 &     0.0810 &    59.8628 &   12054.4  & 1704.83   & 30.14 \\  
SLy   & 11.74   & 0.1766 &     0.0767 &    55.8421 &   11244.8  & 1712.55   & 30.2785 \\  
APR   & 11.37   & 0.1819 &     0.0768 &    48.2159 &    9709.13  & 1790.2   &  31.6514  \\  
FPS   & 10.85   & 0.1907 &     0.0662 &    32.7966 &    6604.17  & 1922.7   & 33.99 \\   
BSK19 & 10.75   & 0.1924 &     0.0647 &    30.6662 &    6175.14  & 1948.14  & 34.444      \\ 
\end{tabular}
\end{ruledtabular}
\end{center}
\end{table*}

In this paper we consider a sample of EOSs taken from the literature.
The sample is chosen to include EOS with a large range of variation
in radius $R$, Love number $k_2$  and tidal parameter $G\mu_2$. 
We consider eleven state-of-the-art EOSs. 
Seven among them, namely, MS1, MS2~\cite{Mueller:1996pm}, 
MPA1, AP3~\cite{PhysRevC.58.1804}, APR, SLy and FPS, have normal 
matter content ($npe\mu$). One, namely GNH3~\cite{Glendenning:1984jr}, 
also incorporates some mixture of hyperons, pion condensates
and quarks. 
Finally, the three labels BSK19, BSK20 and BSK21 refer to 
Skyrme--force--related energy density functionals 
(fitted to nuclear mass data) from which one can compute the
EOS of cold neutron star matter~\cite{Chamel:2011aa}.
Among these equations of state, seven of them (MS1, MS2, MPA1, AP3, SLy, FPS and GNH3)
have been used in Ref.~\cite{Hinderer:2009ca}.
Table~\ref{tab:tableNS} lists, by order of decreasing radius (or increasing compactness) 
the main characteristics of Tolman-Oppenheimer-Volkoff neutron star models built 
from these EOS having mass $1.4M_{\odot}$. These NS properties were computed starting
from the tabulated EOS, using Hermite polynomials interpolation~\cite{Bernuzzi:2008fu}, 
for all EOS but~\hbox{(MS1, MS2, MPA1, AP3)}, that we read instead 
from Table~I of Ref.~\cite{Hinderer:2009ca}.
In our Table~\ref{tab:tableNS}, $\k$ refers to a fiducial $1.4M_\odot + 1.4M_\odot$
binary system and $f^{\rm c}\equiv f^{\rm contact}$ refers to the contact frequency defined above 
(see Eq.~\eqref{eq:x_contact}). Note that $\k$ and $G\mu_2$ decrease
correlatively with the radius due to the dominant influence of the fifth power
of the radius in $\k$ and $G\mu_2$, and in spite of the nonmonotonic behavior 
of $k_2$.
The fifth root of $G\mu_2$ defines a length scale which we can call the {\it tidal radius}
of the NS. It is related to the radius $R$ according to
\be
R_{\rm tidal}\equiv \left(G\mu_2\right)^{1/5}=\left(\dfrac{2}{3}k_2\right)^{1/5} R.
\ee
The values of the tidal radius for the $1.4M_\odot$ models listed in Table~\ref{tab:tableNS}
vary between 8.8195~km (for MS1) and 5.7297~km (for BSK19). The median value is
around 7~km.
In the following we shall focus on a subsample of the EOS listed above, namely
we shall consider only GNH3, BSK21, BSK20, SLy, APR, FPS and BSK19, which span
a plausible subrange of values of $G\mu_2$ (note that we conservatively 
eliminate for instance the very stiff EOS MS1 which yields an extremely 
large value of $G\mu_2$).

\subsection{Measurability of $G\mu_2$: equal-mass case}
\label{sec:mu_numbers}

In this section we focus on the measurability of tidal parameters in {\it equal-mass} 
BNS systems. We shall see below that, within the reasonable range of mass ratios 
expected from observational data, this equal-mass study is a sufficiently accurate
indicator of the general case.

Among the EOSs listed in the previous section, two of them (FPS and BSK19), 
which would have given the two smallest values of $G\mu_2$, lead to maximum 
neutron star masses which are smaller than the recently reported 
value~\cite{Demorest:2010bx} $M_{\rm NS}=(1.97\pm 0.04)M_\odot$.
Because of this we shall first discuss the measurability of tidal parameters within
the restricted, observationally compatible, EOS subsample  GNH3, BSK21, BSK20, SLy, APR.
For each of these EOS we computed the  $5\times 5$ {\it reduced} Fisher matrix $\hat{F}_{ij}$, 
Eq.~\eqref{red_fish}, corresponding  to the parameters 
$[\lambda_1,\dots,\lambda_4;\,\lambda_T]$ where the first four parameters refer 
to the binary system (see Eqs.~\eqref{eq:lambda1}-\eqref{eq:lambda4}), while 
the tidal parameter $\lambda_T$, defined in Eq.~\eqref{eq:def-meanlamb}, 
reduces simply to $G\mu_2=G\mu_2^A=G\mu_2^B$ in the equal-mass case.
The computation of the Fisher matrix elements is performed by considering
that the GW signal is cut off above the (compactness dependent) contact 
frequency, Eq.~\eqref{eq:x_contact}, i.e. each integral is taken over 
the frequency window  $[f_{\rm min},f_{\rm max}]$, with 
$f_{\rm min}=10$~Hz and $f_{\rm max}=f_{\rm contact}$.

The diagonal elements of the inverse of the matrix $\hat{F}_{ij}$ 
yield, according to Eq.~\eqref{sigma_hat}, the SNR-normalized  
probable (statistical) errors $\hat{\sigma}_{\lambda_i}\equiv \rho\sigma_{\lambda_i}$ 
on each parameter $\lambda_i$. 
Before discussing the measurability of the nontidal parameters, let us
start by considering the measurability of the tidal parameter $\lambda_T=G\mu_2$.
\begin{figure}[t]
\center
\includegraphics[width=0.5\textwidth]{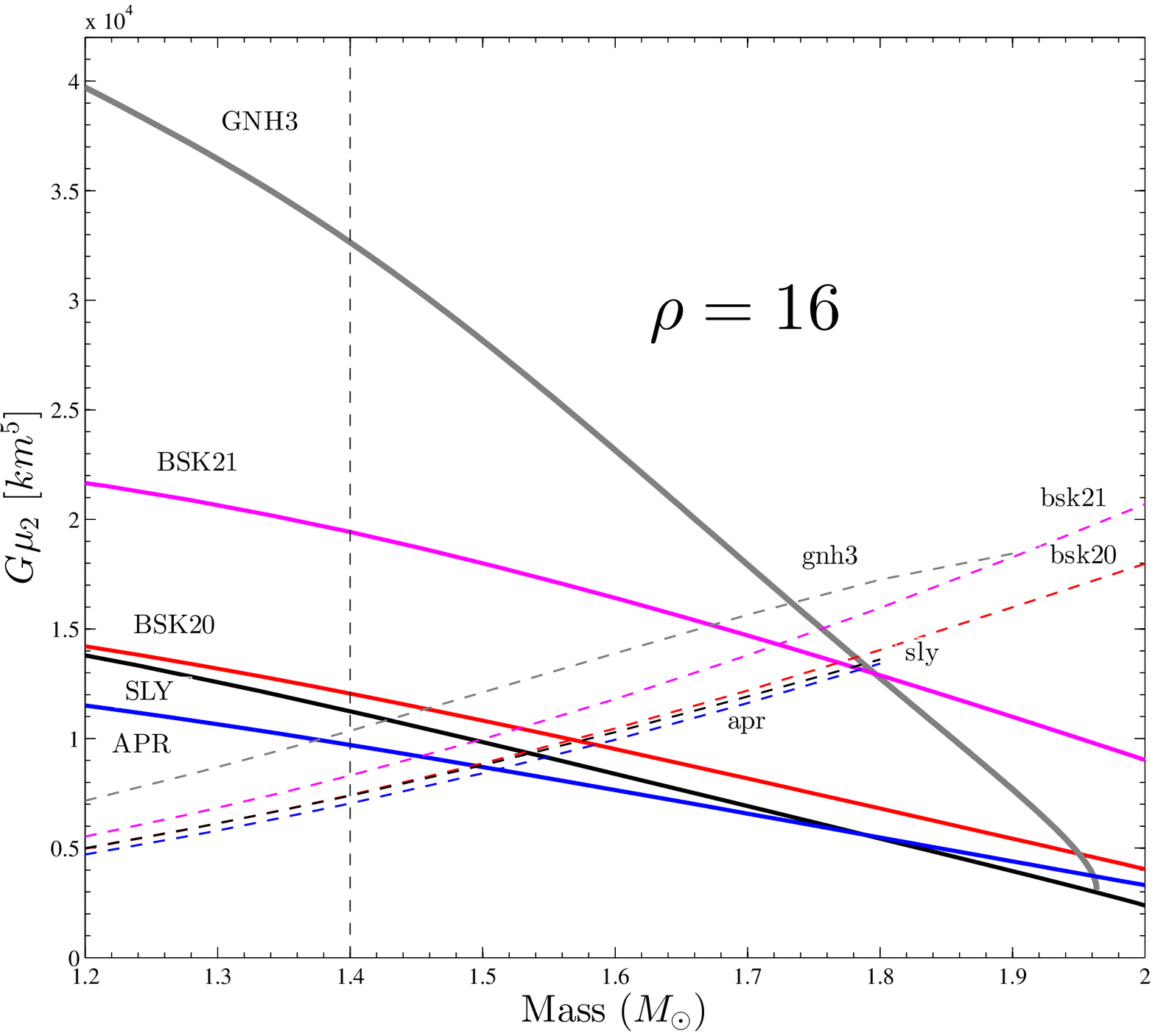}
\caption{\label{fig:measure_mu2}Measurability of the tidal polarizability parameter $G\mu_2$ (in units of km$^5$) 
as a function of the neutron star mass for a sample of realistic EOS from Table~\ref{tab:tableNS}. This plot
refers to the observation (at the SNR level $\rho=16$) of the gravitational wave signal from an equal-mass 
BNS merger as seen by a single advanced LIGO detector. 
The solid lines represent the values of $G\mu_2$ as a function of the NS mass, while the dashed lines represent
the $1~\sigma$ ($68\%$ confidence level) expected statistical errors. The vertical line marks the canonical
NS mass $1.4M_\odot$. Note that over a wide range of masses each solid line lies  comfortably above the corresponding
measurability threshold, therefore indicating that the advanced LIGO-Virgo detector network {\it can} significantly 
 measure $G\mu_2$.}
\end{figure}
\begin{table*}[t]
\caption{\label{table_beta}Measurability of the tidal parameter $G\mu_2$ for a $M=1.4M_\odot + 1.4 M_\odot$ 
neutron star binary obtained using the TaylorF2  frequency-domain approximant to the phase, truncated 
at 2.5PN fractional accuracy for the tidal part and at 2PN accuracy in the point-mass part.
 From left to right the columns report: the name of the EOS; the value of the spin-orbit parameter and of the prior on it; 
the radius of the star; its compactness; the value of the $\ell=2$ tidal parameter $G\mu_2$; the SNR-normalized relative 
errors on the chirp mass $\hat{\sigma}_ {\cal M}/{\cal M}$ and on the $\nu$ parameter $\hat{\sigma}_\nu/\nu$ when cutting 
at bare contact frequency; the SNR-normalized absolute error, $\hat{\sigma}_{G\mu_2}$,
and relative error $\hat{\sigma}_{G\mu_2}/(G\mu_2)$ on $G\mu_2$.
The last two columns refer to the absolute and relative errors on $G\mu_2$ that are obtained 
by taking as cut-off frequency the conservative value 450Hz.}
\begin{center}
\begin{ruledtabular}
\begin{tabular}{ll|lllllll||ll}
 EOS &  $\beta$  & $R$ [km] & ${\cal C}$  & $G\mu_2\,[{\rm km}^5]$ 
                                   & $\hat{\sigma}_{ \ln {\cal M}}$ & $\hat{\sigma}_{\ln \nu}$  & $\hat{\sigma}_{G\mu_2}$ [km$^5$] 
                                   & $\hat{\sigma}_{\ln G\mu_2}$ & $\hat{\sigma}_{G\mu_2}^{450 \rm Hz}$ [km$^5$] & $\hat{\sigma}_{\ln G\mu_2}^{450 \rm Hz}$\\
\hline
\hline
GNH3  & $|\beta|<+\infty$ & 14.19 & 0.1457  &  32641.6  &   0.00415853  & 3.18959  &  186 292 & 5.70720  &  1 476 380 & 45.23   \\
      & $|\beta|<8.5$     & 14.19 & 0.1457  &  32641.6  &   0.00405962  & 3.09906  &  182 612 & 5.59447 &  1 236 580 & 37.8835 \\
      & $|\beta|<0.2$     & 14.19 & 0.1457  &  32641.6  &   0.000447397 & 0.122751 &  165 714 & 5.07679 &    874 001 & 26.7757 \\
      & $\beta=0$         & 14.19 & 0.1457  &  32641.6  &   0.000450135 & 0.117804 &  165 652 & 5.07487 &    873 019 & 26.7456 \\
\hline
BSK21 &  $|\beta|<+\infty$ & 12.57   & 0.1645 &  19424.9  &   0.003946    & 2.98317  &  158 080 & 8.13801   &  1 539 610 & 79.2596  \\
      &  $|\beta|<8.5$     & 12.57   & 0.1645 &  19424.9  &   0.0038749   & 2.91796  &  155 190 & 7.98922 &  1 284 240 & 66.1132  \\
      &  $|\beta|<0.2$     & 12.57   & 0.1645 &  19424.9  &   0.000434397 & 0.115657 &  133 108 & 6.85246 &    876 337 & 45.1141  \\
      &  $\beta=0$         & 12.57   & 0.1645 &  19424.9  &   0.000436901 & 0.110806 &  133 046 & 6.84928 &    875 290 & 45.0603\\
\hline
BSK20 &  $|\beta|<+\infty$ & 11.75   & 0.1760 & 12054.4  &  0.00384331  & 2.88426  &  148 380 & 12.3092  &  1 575 360 & 130.687 \\
      &  $|\beta|<8.5$     & 11.75   & 0.1760 & 12054.4  &  0.00378349  & 2.82927  &  145 750 & 12.0910   &  1 311 380 & 108.788 \\
      &  $|\beta|<0.2$     & 11.75   & 0.1760 & 12054.4  &  0.000428026 & 0.112247 &  118 815 & 9.85656  &    877 640 & 72.8064 \\
      &  $\beta=0$         & 11.75   & 0.1760 & 12054.4  &  0.000430414 & 0.107437 &  118 751 & 9.85125  &    876 558 & 72.7166 \\
\hline
SLy   &  $|\beta|<+\infty$ & 11.74   & 0.1766  &  11244.8   &   0.00383898  &  2.8801   &  148 911 & 13.2426 &  1 579 310 & 140.448  \\
      &  $|\beta|<8.5$     & 11.74   & 0.1760  &  11244.8   &   0.00377961  &  2.82552  &  146 254 & 13.0064 &  1 314 390 & 116.888  \\
      &  $|\beta|<0.2$     & 11.74   & 0.1760  &  11244.8   &   0.000427755 &  0.112104 &  118 271 & 10.5179 &    877 784 &  78.0612 \\
      &  $\beta=0$         & 11.74   & 0.1760  &  11244.8   &   0.000430139 &  0.107295 &  118 206 & 10.5121 &    876 697 &  77.9646 \\
\hline
APR   &  $|\beta|<+\infty$ & 11.37   & 0.1819 & 9709.13 & 0.00379747  & 2.84028  &  142 857 & 14.7136  &  1 586 810 & 163.434  \\
      &  $|\beta|<8.5$     & 11.37   & 0.1819 & 9709.13 & 0.00374226  & 2.78947  &  140 408 & 14.4615  &  1 320 100 & 135.964  \\
      &  $|\beta|<0.2$     & 11.37   & 0.1819 & 9709.13 & 0.000425161 & 0.110728 &  112 643 & 11.6018  &    878 055 &  90.436  \\
      &  $\beta=0$         & 11.37   & 0.1819 & 9709.13 & 0.000427498 & 0.105935 &  112 580 & 11.5953  &    876 961 &  90.3233 \\
\hline
FPS   & $|\beta|<+\infty$  & 10.85  & 0.1907  &  6604.17 & 0.00373437  & 2.77992  &  135 473 & 20.5133  &  1 602 010 & 242.575 \\
      & $|\beta|<8.5$      & 10.85  & 0.1907  &  6604.17 & 0.00368509  & 2.73448  &  133 267 & 20.1792  &  1 331 690 & 201.644 \\
      & $|\beta|<0.2$      & 10.85  & 0.1907  &  6604.17 & 0.000421197 & 0.108641 &  104 424 & 15.8118  &    878 605 & 133.038 \\
      & $\beta=0$          & 10.85  & 0.1907  &  6604.17 & 0.000423462 & 0.103871 &  104 362 & 15.8025  &    877 496 & 132.87  \\
\hline
BSK19 & $|\beta|<+\infty$  & 10.75  & 0.1924 & 6175.14  & 0.00372323  & 2.76928  &  134 005 & 21.7007  &  1 604 110 & 259.769 \\
      & $|\beta|<8.5$      & 10.75  & 0.1924 & 6175.14  & 0.00367495  & 2.72475  &  131 846 & 21.3511  &  1 333 300 & 215.914 \\
      & $|\beta|<0.2$      & 10.75  & 0.1924 & 6175.14  & 0.000420494 & 0.108273 &  102 998 & 16.6795  &    878 681 & 142.293 \\
      & $\beta=0$          & 10.75  & 0.1924 & 6175.14  & 0.000422746 & 0.103507 &  102 937 & 16.6696  &    877 570 & 142.113 \\
\end{tabular}
\end{ruledtabular}
\end{center}
\end{table*}

A recent summary~\cite{Abadie:2010cf} of the expected event rate of BNS coalescences
suggests that at the standard SNR detection threshold $\rho=8$ a ``realistic'' 
estimate of the number of events per year detectable by the advanced LIGO-Virgo network 
is~$\sim 40$. 
This means that at the SNR $\rho=16$ one can reasonably expect to detect $\sim 40/(2^3)=5$ events per year. 
Considering such a SNR $\rho=16$ we plot in Fig.~\ref{fig:measure_mu2}, for each of the 
five EOS selected above, the following two curves: (i) the value of the tidal parameter 
$G\mu_2$ (in $[{\rm km}^5]$) as a function of the mass of each NS (thick, solid lines);
and (ii) the corresponding value of the absolute statistical error $\sigma_{G\mu_2}$ 
(in $[{\rm km}^5]$).
To guide the eye, a vertical line indicates the ``canonical'' mass value $M=1.4M_\odot$.
If we first focus on this mass value, this figure shows that a single advanced LIGO or Virgo
detector {\it can measure} $G\mu_2$ for all considered EOS (with $M_{\max}\geq 1.97M_\odot$)
at a signal to noise ratio $G\mu_2/\sigma_{G\mu_2}$ that varies between 1.4 for the APR EOS 
up to 3.1 for the GNH3 one~\footnote{These measurability ratios refer to observations 
by a {\it single} detector. Observing the same individual BNS event with a network of 3 LIGO-Virgo
detectors will improve the measurability by a factor of order $\sqrt{3}=1.73$, thereby leading to 
signal to noise ratios $G\mu_2/\sigma_{G\mu_2}$ varying between $2.4$ for the APR EOS 
up to 5.4 for the GNH3 one.}.
 
For mass values smaller than $1.4M_\odot$ the measurability of $G\mu_2$ is even better
(larger ratio between $G\mu_2$ and $\sigma_{G\mu_2}$), while for mass values larger than
1.4$M_\odot$ the measurability degrades. The intersection points in Fig.~\ref{fig:measure_mu2}
between solid and dashed curves corresponding to the same EOS mark the value the mass
where $G\mu_2$ is only measurable at the ``$1\sigma$'' ($68\%$ confidence) level, i.e., $\sigma_{G\mu_2}=G\mu_2$.
For instance, for the APR EOS, still assuming a SNR $\rho=16$, equal-mass BNS systems with
individual NS masses larger than $1.52M_\odot$ cannot allow one to extract $G\mu_2$ at a significant level.
By contrast, in the case of BSK21 and GNH3 EOS one can extract tidal parameters for BNS systems
up to individual masses larger than about $1.74M_\odot$.

In summary, Fig.~\ref{fig:measure_mu2} shows that gravitational wave observations from a 
single advanced detector are able
to extract tidal parameters at a significant level, even for the soft EOSs that lead to
the smallest values of $G\mu_2$. This conclusion strikingly contrasts with that of
Hinderer~et~al.~\cite{Hinderer:2009ca}. We will discuss below the reasons behind 
this difference in conclusion.

To complement the graphical representation of our results in Fig.~\ref{fig:measure_mu2},
we present in Table~\ref{table_beta} numerical data referring not only to the 
$5\times 5$, spinless, Fisher matrix calculation behind this figure, but to other
calculations. More precisely, this table gives SNR-normalized errors $\hat{\sigma}_{\lambda_i}$ 
for all parameters of direct physical significance, namely
 $\lambda_3={\cal M}$, $\lambda_4=\nu$ and $\lambda_T=G\mu_2$. [ Note that the numerical 
value of each $\hat{\sigma}_{\lambda_i}$ formally gives the error corresponding to a {\it unit}
 SNR, $\rho=1$. For larger values of $\rho$ the error has to be divided 
by $\rho$.].
This table now considers the larger sample of EOS made by 
GNH3, BSK21, BSK20, SLy, APR, FPS and BSK19. For each one
of these EOS we computed a  $6\times 6$ (or $5\times 5$, see below)
reduced Fisher matrix $\hat{F}_{ij}$, 
Eq.~\eqref{red_fish}, corresponding  now to the parameters 
$[\lambda_1,\dots,\lambda_4;\,\beta ;\,\lambda_T]$.
Here, in addition to the first four binary--system parameters 
considered above and of the tidal parameter $\lambda_T$ we also
consider the spin-orbit parameter $\beta$ (which will be treated
with various different constraints, see below).
We use the same frequency window  
( $[f_{\rm min},f_{\rm max}]$, with $f_{\rm min}=10$~Hz 
and $f_{\rm max}=f_{\rm contact}$) as above.
We now consider the diagonal elements of the inverse of the 
matrix $\hat{F}_{ij}$ for all the parameters of direct physical
significance, namely $\lambda_3={\cal M}$, $\lambda_4=\nu$ and $\lambda_T=G\mu_2$,
we list in Table~\ref{table_beta} the corresponding SNR-normalized
errors $\hat{\sigma}_{\lambda_i}$.

For each EOS, the results are displayed along four rows. On each row, the first
four columns give: (i) information about the treatment of the spin-orbit parameter
$\beta$; (ii) the value of the neutron star radius (in km); (iii) the value of the
compactness; (iv) the value of the tidal parameter $G\mu_2$. The following four
columns give: (v) the fractional, SNR normalized, error on the chirp mass 
$\hat{\sigma}_{\ln {\cal M}}\equiv  \hat{\sigma}_{\cal M}/\M$; (vi) the fractional SNR normalized,
error on the symmetric mass ratio, $\hat{\sigma}_{\ln \nu}\equiv  \hat{\sigma}_\nu/\nu$;
(vii) the {\it absolute}, SNR normalized error $\hat{\sigma}_{G\mu_2}$ on $G\mu_2$ (in $[{\rm km}^5]$); 
and finally (viii) the fractional, SNR normalized error $\hat{\sigma}_{\ln G\mu_2}\equiv \hat{\sigma}_{G\mu_2}/(G\mu_2) $ on $G\mu_2$.
Concerning the treatment of the spin-orbit parameter, the first row, labelled with $|\beta|<+\infty$ refers to
a $6\times 6$ Fisher matrix analysis where $\beta$ is included as a sixth {\it unconstrained} parameter.
The second row, $|\beta|<8.5$, refers to a a $6\times 6$ Fisher matrix analysis where $\beta$ is constrained
by adding a Gaussian prior proportional to $\exp[-1/2(2 \beta/8.5)^2]$. Similarly, the third row corresponds
to a more constraining prior  proportional to $\exp[-1/2(2 \beta/0.2)^2]$. Finally, 
the fourth row corresponds to a $5\times 5$ Fisher matrix analysis where $\beta$ 
is set to zero from the beginning without being fitted for, which was used 
to obtain the data displayed in Fig.~\ref{fig:measure_mu2}.
As already mentioned above, the results for the strong prior $|\beta|<0.2$ (3rd row) 
are nearly indistinguishable from the results of the $5\times 5$ Fisher matrix
analysis (4th row). This justifies our use of the $5\times 5$ Fisher matrix results
in Fig.~\ref{fig:measure_mu2} above.
By contrast, we see that the results corresponding either to the conservative prior
$|\beta|<8.5$ (second row) or the lack of any prior (first row) are close to each other
but differ from the strongly $\beta$-constrained results by very significant factors.
To be precise, the measurability of the chirp mass is worsened by a factor larger than
seven; that of the symmetric mass ratio is worsened by a factor of order 30!; finally,
that of $G\mu_2$ is only worsened by about $20\%$.
These results are linked to the different origins of the effective signals contributing
to the measurability of the various parameters displayed in Fig.~\ref{fig:gamma}.

\begin{table*}[t]
\caption{\label{table_unequal}Measurability of the tidal parameter $\bar{\mu}_2$, 
Eq.~\eqref{eq:def-meanlamb}, for a $M=1.1529M_\odot + 1.6470 M_\odot$ neutron star 
binary ($M_A/M_B=0.7$) obtained using the TaylorF2 frequency-domain approximant at 
2.5PN fractional accuracy in the tidal part of the phase and at 2PN accuracy in 
the point-mass part of the phase. From left to right the columns report: the name 
of the EOS; the compactnesses of the stars; the values of the $\ell=2$ tidal 
polarizability parameters $G\mu_A$ and $G\mu_B$ of the stars; the value of the tidal parameter 
$G\bar{\mu}_2$; the bare contact frequency $f^{\rm c}_{\rm bare}$ in Hz; the SNR-normalized relative 
errors on the chirp mass $\hat{\sigma}_ {\cal M}/{\cal M}$  and on the $\nu$ parameter $\hat{\sigma}_\nu/\nu$; 
the absolute (in km$^5$) and relative errors on $G\bar{\mu}_2$. For each EOS, the second row recalls
the corresponding results for equal masses.}
\begin{center}
\begin{ruledtabular}
\begin{tabular}{lllllllllll}
 EOS &  ${\cal C}_A$ & ${\cal C}_B$  & $G\mu_{2}^A[{\rm km}^5]$ & $G\mu_{2}^B[{\rm km}^5]$ & $G\bar{\mu}_2[{\rm km}^5]$ & $f^{\rm c}_{\rm bare}$ [Hz]  
     & $\hat{\sigma}_{\ln {\cal M}}$  & $\hat{\sigma}_{\ln \nu}$  & $\hat{\sigma}_{G\bar{\mu}_2}$ [km$^5$] & $\hat{\sigma}_{\ln G\bar{\mu}_2}$ \\
\hline
\hline
GNH3       & 0.118   & 0.178  &  41115.1 & 20711.1 & 36178.1 &  1301.97  &   0.000429438 & 0.115405 & 158 647 & 4.38517    \\
GNH3       & 0.1457  & 0.1457 &  32641.6 & 32641.6 & 32641.6 &  1284.06 & 0.000450135 & 0.117804 &  165 652 & 5.07487  \\
\hline
BSK21      & 0.1361  & 0.1938   & 22049.1  & 15623.7 & 21034.5  &  1547.11    &   0.000417681 & 0.108921 & 129 240 & 6.1442 \\
BSK21     & 0.1645   & 0.1645   &  19424.9 & 19424.9 & 19424.9 &  1540.29 & 0.000436901 & 0.110806 &  133 046 & 6.84928 \\
\hline
BSK20      & 0.1446  & 0.2091  & 14638.3   &  8891.9 & 13429.4   &    1713.76    &   0.00041153 & 0.105588 & 115 330 & 8.5879 \\
BSK20      & 0.1760  & 0.1760 & 12054.4   & 12054.4  & 12054.4   & 1704.83      &  0.000430414 & 0.107437 &  118 751 & 9.85125 \\
\hline
SLy        & 0.144   & 0.2116   & 14347.3 & 7696.2   & 12794.1  &  1723.62   &  0.000411201 & 0.105412 & 114 634. & 8.95991 \\
SLy        & 0.1760  & 0.1760  &  11244.8 &  11244.8   & 11244.8   &  1712.55 & 0.000430139 &  0.107295 &  118 206 & 10.5121 \\
\hline
APR        & 0.14934 & 0.2157   &  11874.0 & 7144.9  & 10868.9  &    1797.14    &  0.000408861 & 0.104155 & 109 549 & 10.0792   \\
APR        & 0.1819  & 0.1819 &    9709.13 & 9709.13 & 9709.13 & 1790.2       & 0.000427498 & 0.105935 &  112 580 & 11.5953  \\
\hline
FPS        & 0.154   & 0.2345   & 9443.33 & 3433.1 &  7830.8     &  1956.63    &  0.000404363 & 0.101758 & 100 125 & 12.7862  \\
FPS        & 0.1907  & 0.1907  &  6604.17 &  6604.17 &  6604.17  & 1922.7      & 0.000423462 & 0.103871 &  104 362 & 15.8025 \\
\hline
BSK19      & 0.1553  & 0.2352   & 8866.4  & 3386.96 & 7411.5  &   1973.3    &  0.000403933 & 0.101529 & 99 261 & 13.3929   \\
BSK19      & 0.1924  & 0.1924 & 6175.14 & 6175.14   & 6175.14  & 1948.14   & 0.000422746 & 0.103507 &  102 937 & 16.6696
\end{tabular}
\end{ruledtabular}
\end{center}
\end{table*}

We can roughly summarize the results for the measurability of the nontidal
parameters (in the strongly constrained $\beta$ cases) in the following way: 
\be
\dfrac{\sigma_\M}{\M}\approx \dfrac{4.3\times 10^{-4}}{\rho},
\ee
and
\be
\label{eq_sigma_nu}
\dfrac{\sigma_\nu}{\nu} \approx \dfrac{0.11}{\rho}.
\ee
For instance, when $\rho=10$ this means that the chirp mass is measured 
to a fractional precision of $4\times 10^{-5}$, while the symmetric mass
ratio is measured at a fractional precision of 0.01. 
As usual, the fractional precision on $\M$ is excellent (and has
not been very significantly worsened by the inclusion of the tidal
term, as shown by comparing to the results of Refs.~\cite{Cutler:1994ys,Poisson:1995ef}).
By contrast, the fractional precision on $\nu$ has been significantly worsened
(by a factor of order $1.7$) compared to Refs.~\cite{Cutler:1994ys,Poisson:1995ef}
when fitting for an extra tidal parameter\footnote{Note that when one is fitting 
for the spin parameter $\beta$, the fractional precision of $\nu$ becomes dramatically 
worsened, down to the level $\hat{\sigma}_{\ln\nu}\sim 2.8$. In the case of EOSs 
GNH3 and BSK21 this renders the fractional accuracy on $\nu$ comparable to the fractional
accuracy on $G\mu_2$. In such a case there can be a large difference in the measurability
of $\lambda_T$, Eq.~\eqref{eq:def-meanlamb} versus $\lambda_T'$, 
Eq.~\eqref{eq:lambda_prime}, especially in view of
the correspondingly large correlation between $G\mu_2$ and $\nu$.}.
This worsening in the measurability of $\nu$ might make it difficult to distinguish 
stars with a mass ratio between 0.75 and 1. For instance, if we considered a BNS
with $M_A=1.2M_\odot$, $M_B=1.6M_{\odot}$ (i.e., $M_A/M_B=0.75$) its symmetric mass ratio  
is $\nu\approx 0.2449$, so that $1-4\nu=0.0204$, corresponding to a fractional 
$\delta\nu/\nu\approx 0.02$. Comparing this with the measurement error in $\nu$ for
$\rho=8$, Eq.~\eqref{eq_sigma_nu}, this is only a $2\sigma$-level deviation. 
Actually, this problem may be cured by doing two separate analyses of the GW data,
one using inspiral data only up to a cut-off frequency small enough to be able to neglect
tidal effects (without trying to fit for tidal parameters), which will probably give
a better estimate of the mass ratio. And a separate analysis of the data up to (and possibly
beyond) the merger aimed at extracting EOS--dependent information.

The last two columns of the table exhibit the SNR-normalized absolute and relative errors
on $G\mu_2$ in the case where one uses as upper frequency cut-off $f_{\rm max}=450$~Hz as done
in Ref.~\cite{Hinderer:2009ca,Flanagan:2007ix}.
The use of such a lower cut-off leads to a dramatic worsening (by a factor~$\sim 7$) 
of the measurability of $G\mu_2$
(the origin of this worsening is illustrated in Fig.~\ref{fig:fig2}, 
which includes a line at 450~Hz).

On the other hand, Hinderer~et~al.~\cite{Hinderer:2009ca} computed a SNR-normalized uncertainty
on $G\mu_2$ for the $1.4M_\odot + 1.4M_\odot$ system equal to 
$\hat{\sigma}_{G\mu_2}^{\rm Hinderer}=35\times 19.3\times 0.66743 10^4\,{\rm km}^5=450.84\times 10^4\,{\rm km}^5$
(see second row of their Table~II which corresponds\footnote{We could not reconcile
the statement in Ref.~\cite{Hinderer:2009ca} that they consider a source at a distance of 
100~Mpc, with an amplitude averaged over sky position and relative inclination, with the SNR 35 quoted 
in their Table~II, which, according to Abadie~et~al.~\cite{Abadie:2010cf} seems to correspond to 
an {\it optimally oriented} source at 100~Mpc.} to a SNR $\rho=35$). Considering for example
the SLy EOS, this is a factor 38 larger than the corresponding result in Table~\ref{table_beta}
for our preferred $5$-parameter analysis.
This large factor can be viewed as originating from the product of several subfactors: (i) a factor
of order $(f^c/450~{\rm Hz})^{2.2}=(1704/450)^{2.2}\approx 18.7$ due (according to Eq.~(23) 
of Ref.~\cite{Hinderer:2009ca}) to their use of a cut-off at 450~Hz;
(ii) a factor~$\sim 1.24$ due their use of a conservative prior ($8.5$) on $\beta$; 
iii) a supplementary factor coming from the fact they also fit for the 2PN spin-spin 
parameter $\sigma$ (with a conservative prior), thereby working with seven correlated parameters.

\subsection{Measurability of $G\bar{\mu}_2$: unequal-mass case}

Let us now consider the measurability of tidal parameters in {\it unequal-mass} 
BNS systems. Following Refs.~\cite{Bulik:2003nc,Hinderer:2009ca} we focus on
comparing the measurability of $G\mu_2$ in a system with a large, but plausible, 
mass ratio $M_A/M_B=0.7$ (corresponding to $\nu=0.2422$) to an equal-mass
system. Taking the total mass of the system to be the canonical $M=2.8M_\odot$,
the mass ratio we chose determines $M_A = 1.1529M_\odot$ and $M_B= 1.6470 M_\odot$.
Here we use the $5$-parameter Fisher-matrix analysis with $\beta=0$.
For the same sample of EOS as in Table~\ref{table_beta}, 
Table~\ref{table_unequal} lists  the individual compactnesses, 
the tidal parameters $G\mu_2^{A,B}$, their combination $G\bar{\mu}_2$ and
the SNR-normalized absolute uncertainty on $G\bar{\mu}_2$, $\hat{\sigma}_{G\bar{\mu}_2}$
as well as the relative one  $\hat{\sigma}_{\ln G\bar{\mu}_2}$.
For improved readability of the table, for each EOS we also include the equal-mass
result of Table~\ref{table_beta} in a second row.

As the numbers in this table show, even the large mass ratio 0.7 that we consider does
not influence much the measurability of tidal parameters. In all cases it improves both 
the absolute and the fractional measurability of $G\bar{\mu}_2$ by about~$10-15\%$
(for the EOSs that lead to \hbox{$M_{\rm max}> 1.97M_\odot$}, i.e. excluding FPS and BSK19).

Note that the computations behind the results in Table~\ref{table_beta}-\ref{table_unequal}
have assumed a cut-off frequency which was a function of the individual compactnesses of the
two stars. In an actual GW data analysis situation we won't have an a priori knowledge of these
compactnesses and therefore we will need a way to internally fix the  value of the frequency up
to which the EOB template  can be considered as a reliable description of the observed GW signal.
One can think of several ways in which this could be done.

A first way is to use the fact that, for a given EOS, the contact frequency $f^{\rm contact}$, or
the contact frequency parameter $x_{\rm contact}$ given by Eq.~\eqref{eq:x_contact} is symmetric
under the $A\leftrightarrow B$ exchange, and therefore it can (in principle) be considered as
a function of $G\bar{\mu}_2$, $M$, and $\nu$ (which are all $A\leftrightarrow B$-symmetric functions).
Moreover, in view of the strong, approximately universal, dependence of $G\mu_2^A$ 
on $\C^A$ ($G\mu_2^A\propto k_2^A(\C_A)\C_A^{-5}$, with $k_2^A(\C_A)\approx A-B\C_A$, 
as per Eq.~(116) of~\cite{Damour:2009vw}, using for $(A,B)$ values appropriate for an 
``average'' EOS) the function determining $f^{\rm contact}$ in terms of $G\bar{\mu}_2$, $M$ and $\nu$
can be considered as approximately universal and known.
A second way is, separately from a tidal--parameter--fitting data analysis of the inspiral signal, 
to use the full GW data including the post-merger signal to extract information both about
the frequency of merger and the post-merger dynamics, so as to have some independent 
handle on the EOS. Indeed, recent numerical results~\cite{Baiotti:2008ra,Bernuzzi:2010xj,Hotokezaka:2011dh,Bauswein:2011tp}
on BNS merger have shown that the GW signal contains definite imprints both of the merger
and post-merger dynamics. For instance, on Fig.~2 of Ref.~\cite{Bauswein:2011tp} both
the frequency marking the end of inspiral (corresponding to $f^{\rm contact}$ in our EOB setup),
and the characteristic frequency of post-merger oscillations stand out above the advanced LIGO
noise.

\subsection{Correlations}
\label{sec:correlations}
\begin{table}[t]
  \caption{\label{tab:Cij} Correlations, as given by Eq.~\eqref{eq:Corr}, 
  between the dynamically relevant parameters, $\{\M,\nu,G\mu_2\}$ of 
  the $1.4M_\odot+1.4M_\odot$ binaries of Table~\ref{table_beta} as obtained from the
   $5\times 5$ Fisher matrix analysis (the spin parameters, $(\beta,\sigma)$, are
  set to zero from the start). From left to right the columns report: the name 
  of the EOS, the cut-off frequency (either contact frequency or 450~Hz), 
  the correlation between  $(G\mu_2,\M)$, the correlation between $(G\mu_2,\nu)$ 
  and the correlation between $(\M,\nu)$. The correlations between 
  $G\mu_2$ and $(\M,\nu)$ decrease when the cutoff frequency is increased.}
\begin{center}
\begin{ruledtabular}
\begin{tabular}{llccc}
 EOS   & $f^{\rm max}$~[Hz]  &$C^{(\M,G\mu_2)}_{35}$ &  $C^{(\nu,G\mu_2)}_{45}$  & $C^{(\M,\nu)}_{34}$  \\
\hline
\hline
 GNH3  &  1284.06 & 0.568314  & 0.755652 & 0.911907 \\
 GNH3  &  450     & 0.693511  & 0.859695 & 0.93168\\
 \hline
 BSK21 & 1540.29  & 0.551751 & 0.73937  & 0.909245\\
 BSK21 & 450      & 0.694515 & 0.860478 & 0.93168\\
\hline
 BSK20 & 1704.83 & 0.543526 & 0.730951 & 0.907868\\
 BSK20 & 450     & 0.695072 & 0.860911 & 0.93168\\
\hline
 SLy   & 1712.55 &0.543398 & 0.730791 & 0.907809\\
 SLy   & 450     & 0.695133 & 0.860959 & 0.93168\\
\hline
 APR  & 1790.2       & 0.539288 & 0.726584 & 0.907232\\
 APR  & 450     & 0.695249 & 0.861049 & 0.93168\\
\hline
 FPS  & 1922.7 & 0.539288 & 0.726584 & 0.907232\\
 FPS  & 450 & 0.695249 & 0.861049 & 0.93168\\
\hline
 BSK19& 1948.14 & 0.532037 & 0.719014 & 0.906174\\
 BSK19& 450     & 0.695515 & 0.861256 & 0.93168
\end{tabular}
\end{ruledtabular}
\end{center}
\end{table}
\begin{table}[t]
  \caption{\label{tab:Gi} The quantities $G_i$, as defined by Eq.~\eqref{eq:Gi}, 
  for the physically relevant parameters, $\{\M,\nu,G\mu_2\}$ of the $1.4M_\odot+1.4M_\odot$ 
  binaries of Table~\ref{table_beta}. These values are obtained from the
   $5\times 5$ Fisher matrix analysis, where the spin parameters, $(\beta,\sigma)$, are
  set to zero from the start. From left to right the columns report: the name 
  of the EOS, the cut-off frequency (either contact frequency or 450~Hz),
  and the $G_i$'s. The value of $G_i$ decreases when the the cutoff frequency 
  is increased.}
\begin{center}
\begin{ruledtabular}
\begin{tabular}{llccc}
 EOS   & $f^{\rm max}$~[Hz]  &$G_3^{(\M)}$ &  $G_4^{(\nu)}$  & $G_5^{(G\mu_2)}$  \\
\hline
\hline
 GNH3  &  1284.06 &  3.80973 & 12.7126 & 4.94471 \\
 GNH3  &  450     &  5.04097 & 21.973  & 10.9479 \\
 \hline
 BSK21 & 1540.29  &  3.69651 & 11.959  &  4.44276 \\
 BSK21 & 450      &  5.04099 & 21.982  & 10.9764  \\
\hline
 BSK20 & 1704.83 &  3.6412  & 11.5971 &  4.20174 \\
 BSK20 & 450     &  5.041   & 21.987  & 10.9923  \\
\hline
 SLy   & 1712.55 & 3.63886 & 11.5821 &  4.19293 \\
 SLy   & 450     & 5.04101 & 21.9876 & 10.994   \\
\hline
 APR  & 1790.2  & 3.57594 & 11.1741 &  3.91152  \\
 APR  & 450     & 5.04101 & 21.9886 & 10.9973   \\
\hline
 FPS  & 1922.7 & 3.58202 & 11.2133  &  3.9388  \\
 FPS  & 450    & 5.04101 & 21.9908  & 11.004   \\
\hline
 BSK19& 1948.14 &  3.57594 & 11.1741 &  3.91152 \\
 BSK19& 450     & 5.04102 & 21.9911  & 11.0049 
\end{tabular}
\end{ruledtabular}
\end{center}
\end{table}
To complement the results about the measurability of $G\mu_2$ (and $G\bar{\mu}_2$) given in 
the previous two sections, let us discuss the issue of the correlations among 
the various parameters and their influence on the measurability of  $G\mu_2$.
Usually, correlations are measured via the nondiagonal terms of the covariance matrix, 
that is by
\be
\label{eq:Corr}
C_{ij}=\dfrac{(\hat{F}^{-1})^{ij}}{\sqrt{(\hat{F}^{-1})^{ii}(\hat{F}^{-1})^{jj}}},
\ee
which are numbers that vary between $-1$ and $+1$.
We focus on canonical  $1.4M_{\odot}+1.4M_{\odot}$  equal-mass binaries built 
from our sample of EOSs.
The values of the correlations (for the dynamical parameters $\{\M,\nu,G\mu_2\}$) 
$C_{ij}$ for the $5\times5$ Fisher matrix analysis are listed in  Table~\ref{tab:Cij}.
For each EOS, we list the values of $C_{ij}$ when one takes as cutoff 
frequency the contact frequency (top row) as well as their values
when taking 450~Hz as cutoff frequency (bottom row).
Note first that $G\mu_2$ is only (especially when using the contact frequency as cutoff) 
moderately correlated to $\M$ and $\nu$: by contrast  to the $(\M,\nu)$ correlation, 
the $(\M,G\mu_2$) and $(\nu,G\mu_2)$ correlations are always comfortably smaller than $0.9$.
For a given cut-off frequency, the values of the correlations $C_{ij}$
decrease when the compactness of the model increases. This decrease
is mild. To be precise, considering the variability between GNH3 
and BSK19, we have that $C^{(\M,G\mu_2)}_{35}$ varies by $6.4\%$,
 $C^{(\nu,G\mu_2)}_{45}$ by $4.8\%$ and $C^{(\M,\nu)}_{34}$ by $0.6\%$.
On the other hand, the values of the correlation {\it increase} 
when the cutoff frequency is {\it decreased} from the contact frequency to 450~Hz.
This is expected since up to 450~Hz the tidal part of the phasing 
is quite weak and thus rather difficult to disentangle from the nontidal signal.
Note that $G\mu_2$ is only moderately correlated to $\M$ and $\nu$: by contrast 
to the $(\M,\nu)$ correlation, the $(\M,G\mu_2$) and $(\nu,G\mu_2)$ correlations
are always comfortably smaller than $0.9$.

It is also useful to look at the quantity 
(for each parameter $\lambda_i$)
\be
\label{eq:Gi}
G_i=\sqrt{(\hat{F}^{-1})^{ii}\hat{F}_{ii}}.
\ee
For each EOS we list the values of $G_i$ (for $i=3,4,5$, i.e. $\M$, $\nu$ and $G\mu_2$) 
in Table~\ref{tab:Gi}.
The quantity $G_i$ measures the {\it global} correlation coefficient, say $c_i^{\rm global}$, 
of the parameters $\lambda_i$ with respect to all other parameters $\lambda_j$, $j\neq i$, 
via
\be
G_i = \dfrac{1}{\sqrt{1-\left(c_i^{\rm global}\right)^2}}.
\ee
Here, $c_i^{\rm global}$ is the larger possible correlation between $\lambda_i$ and
a linear combination of the other parameters $\lambda_j$, $j\neq i$.
Let us discuss in more detail the meaning of the quantities $G_i$.  We recall
that it is convenient to interpret the measurability of the various 
parameters $\lambda_i$ entering the
phasing signal $\Psi(\f;\lambda_i)$ in terms of geometrical concepts
related to the
scalar product  \eqref{renorm_prod} (which is the SPA version of the
Wiener scalar product).
When considering small variations $\delta \lambda_i$
of the parameters, to each parameter  $\lambda_i$
is associated the vector $\Psi_i \equiv \de_{\lambda_i}\Psi$ so that the
infinitesimal
signal associated to a simultaneous variation of all the parameters is the
following
linear combination of  individual vectors: $\delta \Psi=\sum_i \delta
\lambda_i \Psi_i$.
The geometrical transcription of the fact that the measurement of a
particular parameter
$\lambda_i$ is
correlated to the measurement of the other parameters $\lambda_j$, 
$j \neq i$, is that the signal vector $\Psi_i$ associated to 
$\delta \lambda_i$ is not orthogonal to the other signal vectors $\Psi_j$. 
[Remember that the Fisher matrix is the matrix
of scalar products $\hat{F}_{ij}= (\Psi_i | \Psi_j)$.]  
The global correlation between $\lambda_i$ and all the other  
$\lambda_j$'s, $j \neq i$, is then measured by the ``inclination angle''
$\theta_i$ between the vector $\Psi_i$ and the hyperplane $H_i$ spanned by
the remaining vectors  $\Psi_j$, $j \neq i$. The angle $\theta_i$ is
defined so that it vanishes when $\Psi_i$ lies within the hyperplane $H_i$, 
and equals $\pi/2$ when $\Psi_i$ is orthogonal to the hyperplane $H_i$.  
Let us now decompose the vector $\Psi_i$
in two orthogonal vectors: (i) its projection $\Psi_i^\perp $ orthogonal
to $H_i$,
and (ii) its projection $\Psi_i^\parallel$ parallel to $H_i$. 
It is then easy to see that the definition of $G_i$ given by Eq.~\eqref{eq:Gi} 
implies
\be
\sin \theta_i = \dfrac{1}{G_i} =\dfrac{| \Psi_i^\perp |}{| \Psi_i |},
\ee
where $| v |$ denotes the (Euclidean) length of the vector $v$ 
in signal space. Note also that the global correlation coefficient 
$c_i^{\rm global}$ defined above is simply equal to
$c_i^{\rm global} =  \cos \theta_i$.

Let us also note the following formulas yielding the SNR-normalized
(absolute and fractional) error(s) on the parameter $\lambda_i$
\be
\hat{\sigma}_{\lambda_i}\equiv\rho\,\sigma_{\lambda_i} = \dfrac{1}{|\Psi_i^\perp |}=\dfrac{G_i}{|\Psi_i|},\\
\ee
\be
\hat{\sigma}_{\ln \lambda_i}                         = 
                          \dfrac{\hat{\sigma}_{\lambda_i}}{\lambda_i}=\dfrac{1}{\lambda_i|\Psi_i^\perp|}=\dfrac{G_i}{\lambda_i |\Psi_i|}.
\ee
In particular, if we apply the last formula to the tidal parameter
$\lambda_T\equiv G\mu_2$ we see that, given
a certain SNR $\rho$, the two factors that determine the measurability of 
$\lambda_T$ are $G_T\equiv G_5^{(G\mu_2)}$ (which measures the adverse effect of correlations 
with the other parameters, and which should be as small as possible), 
and the Euclidean length of the full tidal signal $| \lambda_T \Psi_T|$. 
In other words,  the ``useful'' part of the explicit frequency-domain 
tidal signal $ \lambda_T \Psi_T(f)$ pictured in Fig.~1 (which reaches
about ten radians at contact) is reduced by two factors: a first factor
coming from the overlap between the SNR measure $\gamma(f)$ and the 
tidal signal $\Psi_T(f) \sim f^{5/3}$
(which enters the integral $| \lambda_T \Psi_T|=\sqrt{\int df \gamma(f)
(\lambda_T \Psi_T(f))^2)}$), and a second factor $1/G_T$,  due to the 
correlations (which retain only the part of the tidal vector which 
is orthogonal to all the other signal vectors).
This motivates us to define the {\it useful number} of radians 
(in a rms sense) contained in the tidal dephasing signal as
\be
\label{eq:psi_useful}
\Psi_T^{\rm useful} \equiv \lambda_T |\Psi_T^\perp| =
\dfrac{\lambda_T |\Psi_T|}{G_T} =\dfrac{1}{\hat{\sigma}_{\ln G\mu_2} }.
\ee
In view of the results reported in Tables~II and~III  a median estimate
for the useful tidal dephasing is $ \Psi_T^{\rm useful} \sim 0.1$ rads. 
This is a factor $\sim 100$ smaller than the dephasing at contact. 
This reduction factor can be seen as the product of a factor $G_T \sim 5$ 
due to correlations, and a factor $\sim 20$ coming from the fact that the 
tidal signal is strongest during late inspiral, when the SNR curve 
$f\gamma(f)$ is much below its maximum (see Fig.~3).  Note also that 
if one uses 450 Hz as cut-off frequency the useful tidal dephasing is 
drastically reduced (roughly by a factor 7): e.g. for the BSK21 EOS
 which led to a rather comfortably measurable tidal signal 
$ \Psi_{T \, {\rm BSK21}}^{\rm useful} \sim 1/7$ when considered up to contact, 
one has only $ \Psi_{T  \, {\rm  BSK21}}^{{\rm useful} 450} \sim 1/45$.  This loss 
in measurability by a factor $\sim 6.6$ is due both to a higher global 
correlation ($G_T$ increasing from 4.44 to 11.0) and to a smaller signal 
at 450~Hz versus $f^c$.
As in the case of the correlations $C_{ij}$, the main message of Table~\ref{tab:Gi}
is that (especially when considering as cutoff frequency the contact frequency) 
the global correlation of $G\mu_2$ with respect to all other parameters is moderate
and comparable to that of $\M$. By contrast, $\nu$ is more strongly correlated
to the other parameters.

\subsection{Coherent data analysis of tidal parameters}

Until now we have been discussing the measurability of tidal parameters from the GW signal
emitted by a {\it single}, particular BNS merger event (eventually simultaneously observed by
3 separate detectors).
We wish now to introduce a new way of extracting EOS-dependent information by a 
``coherent'' data analysis of the GW signals emitted by many separate BNS merger events,
say the expected~$\sim 40$ BNS mergers observable in one year by one advanced LIGO (or Virgo)
detector. [Evidently, the method can also be extended to a coherent analysis of the data
coming from the full network of LIGO-Virgo detectors].

This method is based on the following preliminary remark. As exemplified on 
our Fig.~\ref{fig:measure_mu2} above (as well as in Fig.~2 of Ref.~\cite{Hinderer:2009ca}),
the tidal parameter $G\mu_2^A$ is, for a given EOS, a function of the mass $M_A$ of the considered 
neutron star which can be well represented by a {\it linear function} in the range of expected
neutron-star masses, $1.2M_\odot-1.9M_\odot$, say
\be
\label{eq:mu2_lin}
G\mu_2^A(M_A) = a_{\rm EOS} + b_{\rm EOS} M_A.
\ee
The crucial point here is that the coefficients $(a_{\rm EOS},b_{\rm EOS})$ depend {\it only} on the EOS,
but not anymore on the specific neutron star mass. In the following, we shall use the symbols
$(a_{\rm EOS},b_{\rm EOS})$ to denote the two unknown parameters  corresponding to the actual EOS chosen by Nature.
Moreover, the system tidal parameter $G\bar{\mu}_2$, Eq.~\eqref{eq:def-meanlamb}, that enters the inspiral 
signal of an individual system $(M_A,M_B)$, becomes, when using Eq.~\eqref{eq:mu2_lin}
\begin{align}
G\bar{\mu}_2&=\dfrac{1}{13}a_{\rm EOS}\left(\dfrac{6}{\nu}-11\right) + \dfrac{1}{2} b_{\rm EOS} M \nonumber\\
\label{eq:mubar_ab}
           &=\dfrac{1}{13}a_{\rm EOS}\left(\dfrac{6}{\lambda_4}-11\right) 
            + \dfrac{1}{2} b_{\rm EOS} \dfrac{\lambda_3}{\lambda_4^{3/5}}. 
\end{align}
Let $N$ denote the number of BNS merger events observed during a certain period $T$ (e.g, 1~yr).
We introduce an index $I=1,\dots,N$ labelling each BNS system, and the corresponding merger event, 
within this collection of observed GW signals.
We now discuss a data analysis procedure for the combined event consisting of this collection 
of $N$ individual GW signals. This ``grand signal'' depends on a collection of parameters:
 $(\lambda_a^I,\,a_{\rm EOS},\,b_{\rm EOS})$. Here the $\lambda_a^I$'s vary from BNS system to BNS system
(and include, besides the parameters $\lambda_{1,\dots,4}$ considered above, also 
an amplitude parameter), while the two EOS parameters, $(a_{\rm EOS},\, b_{\rm EOS})$ are 
common to the whole collection of events.
We now envisage a grand fit of the whole collection of parameters $(\lambda_a^I,\,a_{\rm EOS},\,b_{\rm EOS})$
to the ensemble of $N$ GW signals. 
The signals in this  ensemble have clearly statistically independent noise contributions.
Let us then consider the Bayesian probability distribution function for the values of the 
parameters $\left(\lambda_a^I,a_{\rm EOB},b_{\rm EOS}\right)$, given a grand strain signal $\{s^I(t)\}$.
It can be written as $\exp\left(-\chi^2/2\right)$ where~\cite{Cutler:1994ys}
\begin{align}
&\chi^2\left(\lambda_a^I,a_{\rm EOS},b_{\rm EOS};\,s^I,{\rm priors}\right)= \sum_{I=1}^N\chi^2_{I\,{\rm prior}}\nonumber\\
 &+\sum_{I=1}^N\langle h\left(\lambda_a^I,a_{\rm EOS},b_{\rm EOS}\right)-s^I, h\left( \lambda_a^I,a_{\rm EOS},b_{\rm EOS}\right)-s^I  \rangle,
\end{align}
with $\langle\dots,\dots\rangle$ denoting as above the single-observation Wiener scalar product, and $\chi^2_{I\,{\rm prior}}$
indicating the logarithms of eventual priors on some parameters (e.g.,$\chi^2_{I\,{\rm prior}}=\left(2\beta_I/\beta_{\rm max}\right)^2$
as above).
For simplicity, let us use only strong priors, that are equivalent to eliminating some parameters 
(e.g., $\beta_{\rm max}\to 0$ equivalent to setting $\beta_I$ to zero). 
In the high SNR approximation, and after having marginalized
over the $N$ amplitude parameters $A_I$ (treated in the Gaussian approximation, and as approximately
independent of the other parameters), we can approximately reexpress $\chi^2$ in terms of phase differences,
using the renormalized scalar product~\eqref{renorm_prod}:
\begin{widetext}
\begin{align}
\chi^2\left(\lambda_a^I,a_{\rm EOS},b_{\rm EOS};\,s^I\right)\simeq \sum_{I=1}^N\rho_I^2
\biggl(\Psi(f;\lambda_a^I,a_{\rm EOS},b_{\rm EOS})-\Psi_{s^I}(f)|\Psi(f;\lambda_a^I,a_{\rm EOS},b_{\rm EOS})-\Psi_{s^I}(f)\biggr)_I.
\end{align}
\end{widetext}
The scalar product~\eqref{renorm_prod} a priori depends on the index $I$ through the choice of the cut-off
frequency $f_{\rm max}=f^c\left(M^I,\C_A^I,X_A^I\right)$. As a first approximation for understanding
how using such a $\chi^2$ improves the measurement of $\left(a_{\rm EOS},b_{\rm EOS}\right)$, let us 
however consider that one uses some a priori fixed cut-off frequency. 
The theoretical phase $\Psi\left(f;\,\lambda_a^I,a_{\rm EOS},b_{\rm EOS}\right)$ depends on $(a_{\rm EOS},b_{\rm EOS})$
only through a term of the form $\lambda_T^I\Psi_{\rm tidal}(f;\,\lambda_a^I)$, where 
$\lambda_T^I=G\bar{\mu}_2^I$ is linear in $a_{\rm EOS}$ and $b_{\rm EOS}$, see Eq.~\eqref{eq:mubar_ab}.
If we then project $\Psi_{\rm tidal}(f)$ into its projection $\Psi_{{\rm tid}\,I}^\perp(f)$, which is orthogonal
to the nontidal signals $\Psi_{\lambda_a^I}(f)\equiv \de_{\lambda_a^I}\Psi^0(f;\lambda_a^I)$,
[with respect to the Euclidean metric $(\dots|\dots)$], we find that the part of $\chi^2$ 
which depends on $(a_{\rm EOS},\,b_{\rm EOS})$ is quadratic in them and of the form
\begin{align}
&\chi^2\left(a_{\rm EOS},b_{\rm EOS}\right)\simeq \sum_{I=1}^N\rho_I^2\left(\Psi_{{\rm tid}\,I}^\perp|\Psi^\perp_{{\rm tid}\,I}\right)\nonumber\\
&\times \left[\dfrac{1}{13}\left(\dfrac{6}{\lambda_4^I}-11\right)a_{\rm EOS}
+\dfrac{\lambda_3^I}{2(\lambda_4^I)^{3/5}}b_{\rm EOS}-c(s^I,\lambda_a^I)\right]^2.
\end{align}
Here (as discussed in the previous section) 
the factor $\left(\Psi_{{\rm tid}\,I}^\perp|\Psi^\perp_{{\rm tid}\,I}\right)$  takes into account the correlation of 
$(a_{\rm EOS},\,b_{\rm EOS})$ (via $\lambda_T^I$) with the nontidal parameters.
Finally, the latter formula defines, for each confidence level, an error ellipse in the 
$(a_{\rm EOS},\,b_{\rm EOS})$ plane. The size of the minor axis of the ellipse (associated to some
best determined $\lambda_T$-like combination of $a_{\rm EOS}$ and $b_{\rm EOS}$) will essentially
be determined by the following effective squared SNR, given by the sum of all individual SNR's, i.e.
\be
\label{eq:SNReff}
\rho^2_{\rm coherent}=\sum_{I=1}^N\rho_I^2.
\ee
On the other hand the size of the major axis of the error ellipse will crucially depend on the
dispersion of the distribution of $\lambda_3=\M$ and $\lambda_4=\nu$ around their median values.
If such a dispersion is large enough, one will be able to measure both $a_{\rm EOS}$ and $b_{\rm EOS}$
with a precision still mainly determined by the effective SNR~\eqref{eq:SNReff}.
If, on the contrary, all observed BNS systems happen to be close to, say, the canonical
$(1.4M_\odot,1.4M_\odot)$ system, the error ellipse will be very elongated in one
direction, while the other direction (minor axis) will lead to the measurement of one $\lambda_T$-like
combination of $a_{\rm EOS}$ and $b_{\rm EOS}$. In all cases, the crucial quantity determining the 
measurability of some $\lambda_T$-like quantity (or quantities) is the effective squared SNR, Eq.~\eqref{eq:SNReff}.
One can then estimate the sum giving the effective squared SNR Eq.~\eqref{eq:SNReff} 
by approximating it by an integral over the ball of space around the
Earth containing all the source events up to the minimum threshold of detectability of a BNS, 
that we shall take as $\rho_{\rm min}=8$. This leads to the following estimate of the coherent SNR
\be
\rho_{\rm coherent}=\sqrt{3N}\rho_{\rm min},
\ee
 where the fact that the $\rho_{\rm min}$ is augmented even beyond the naively expected $\sqrt{N}$ factor
is due to the existence of closer events having a larger SNR.
If we apply this result to an expected ``realistic'' number of events, $N\sim 40$ during one year of
observation, we conclude that the effective SNR for such a coherent analysis 
is $\sim \sqrt{3\times 40}\times 8\approx 88$. On the other hand, there is clearly a price to
pay for such an remarkable increase in SNR. Indeed, as sketched above, the two parameters $(a_{\rm EOS},\, b_{\rm EOS})$ 
will now be correlated among themselves, which will degrade their individual measurability.
We leave to future work a detailed exploration of the performance of such a coherent analysis
(taking into account all the correlations that have been neglected in the sketchy treatment above),
especially for the measurability of  $(a_{\rm EOS},\, b_{\rm EOS})$. We however expect that it will
significantly improve the single detector measurability computed above, thereby strongly 
reinforcing our conclusion that the advanced LIGO-Virgo network can extract EOS information
from the late inspiral BNS signal.

\subsection{Sensitivity to $\alpha_2^{(2)}$}

In the analysis of this paper we have used the recently analytically computed value of 
the 2PN tidal amplification parameter, Eq.~\eqref{eq:alpha2}, leading to 
$\alpha_2^{(2)}=85/14\approx 6$ in the equal-mass case.
However, the comparison between the waveform prediction coming from the EOB tidal model
and recent, state-of-the-art BNS numerical simulations~\cite{Baiotti:2010xh,Baiotti:2011am} 
has suggested that the effective value of $\alpha_2^{(2)}$ might be, in the equal-mass case, 
of order~40 or even larger. In addition, analytical arguments have been advanced in 
Ref.~\cite{Bini:2012gu} suggesting  that $\alpha_2^{(2)}$ might be further amplified by higher
PN effects, possibly by a factor of order 2.
In view of this uncertainty on the influence of higher relativistic corrections to the 
tidal interaction energy, we have explored the effect on the measurability of $G\mu_2$ 
of changing the value of $\alpha_2^{(2)}$.
Our results are displayed in Table~\ref{tab:vary_alpha2}, which for a large sample of 
EOS lists the value of the SNR-normalized fractional error  $\hat{\sigma}_{\ln G \mu_2}$ on $G\mu_2$
for canonical $1.4M_\odot+ 1.4M_\odot$ systems. As we see from the numbers in the table,
if it turns out that the effective value of $\alpha_2^{(2)}$ is closer to 40 than to 6, this
will improve the measurability of $G\mu_2$ by more than $20\%$. Therefore, all our conclusions
above should be considered as conservative from this point of view.
Note also that even using the value $\alpha_2^{(2)}=0$ does not degrade by more than $5\%$ the
measurability of $G\mu_2$. This is due to the presence of a rather large contribution (6.99)
in the coefficient of $x^2$ in Eq.~\eqref{eq:2.5PNhat} giving the PN correction factor to the
tidal phase.
\begin{table}[t]
  \caption{\label{tab:vary_alpha2} Sensitivity of $\hat{\sigma}_{\ln G \mu_2}$ when varying $\alpha_2^{(2)}$.}
\begin{center}
\begin{ruledtabular}
\begin{tabular}{lccc}
 EOS  & $\hat{\sigma}_{\ln G \mu_2}^{\alpha_2^{(2)}=0} $  & $\hat{\sigma}_{\ln G \mu_2}^{\alpha_2^{(2)}=85/14}$ &  $\hat{\sigma}_{\ln G \mu_2}^{\alpha_2^{(2)}=40}$ \\
\hline
MS1   &    3.5010 &     3.3733 &       2.7975 \\
GNH3  &    5.2783 &     5.0749  &      4.1688 \\ 
MS2   &    5.7074 &     5.4774  &      4.4629 \\
BSK21 &    7.1710 &     6.8493  &      5.4649 \\
MPA1  &    7.4385 &     7.1011  &      5.6529 \\
AP3   &    8.8549 &     8.4382  &      6.6663 \\
BSK20 &   10.3566 &     9.8513  &      7.7224 \\
SLy   &   11.0534 &    10.5121  &      8.2342 \\
APR   &   12.2162 &    11.5953  &      9.0079 \\
FPS   &   16.7035 &    15.8025  &     12.1087 \\
BSK19 &   17.6312 &    16.6696  &     12.7398 
\end{tabular}
\end{ruledtabular}
\end{center}
\end{table}

\section{Conclusions}
\label{sec:end}

To conclude, let us summarize the main steps of our analysis as well as our main results.
\begin{itemize}

\item For describing the non-tidal, point-mass contribution to the Fourier domain phasing, $\Psi^0(f)$, 
we used a 2PN TaylorF2 approximant, which suffices for our purposes given the fact that the 
measurability of the {\it nontidal} parameters $(\M,\,\nu)$ is mainly contained in a 
rather early part of the inspiral signal, say for frequencies $f\lesssim 50$~Hz (see Fig.~3).

\item By contrast, as the tidal contribution to the Fourier domain phasing $\Psi^T(f)$ 
is dominated by the late-inspiral part of the waveform, 
we had to work harder and to use the  tidal-EOB formalism as a way to define a controlled, 
analytical description of the phasing of tidally interacting BNS systems up to merger.
The controlled analytical description we use is a mixture of stationary phase approximation
and of suitably accurate post-Newtonian expansions.
We have checked that our approximation to the tidal EOB dephasing is accurate to better than
0.3~rad up to merger.

\item The final total phasing $\Psi(f)=\Psi^0(f)+\Psi^T(f)$ depends on a number of parameters.
We first used observational data of known binary pulsars to set an apriori upper bound on the 
magnitude of the spin-dependent effects of $\Psi^0(f)$. These effects are proportional to a
spin-orbit parameter, $\beta$ and a spin-spin one, $\sigma$. We found that observational data
suggest that $|\beta|<0.2$ and $|\sigma|\lesssim 10^{-4}$. We showed that enforcing these upper
bounds as Bayesian priors in a data analysis is essentially equivalent to neglecting from the
start $\beta$ and $\sigma$. 

\item We have computed $G\mu_2$ for a large sample of EOSs (including three recently defined 
EOSs: SK19, BSK20 and BSK21~\cite{Chamel:2011aa}).

\item The previous considerations allowed us to perform a data analysis based on a 
Fisher matrix formalism containing five parameters $\{t_c,\phi_c,\M,\nu,G\mu_2\}$, and
taking into account an EOB-controlled analytical GW signal going up to merger.

\item The main result of our analysis is that the tidal polarizability coefficient $G\mu_2$ 
{\it can be measured} at the $95\%$ confidence level by the advanced LIGO-Virgo detector network 
using GW signals with reasonable SNR ($\rho=16$). This measurability result holds for all the
EOSs in the sample we have considered, under the only restriction that their maximum mass be 
larger than the recently observed NS mass $1.97M_\odot$~\cite{Demorest:2010bx}.  
This measurability property is true for BNS mass ratios at least between 0.7 and 1.

\item We proposed a promising new way of extracting EOS-dependent information from the coherent analysis
of a collection of GW observations of separate BNS merger events. 

\item The latter method is based on parametrizing the unknown EOS-dependent {\it function}
$G\mu_2(M_A)$ by a (local) linear fit, Eq.~\eqref{eq:mu2_lin}, depending on only two
parameters: $(a_{\rm EOS},b_{\rm EOS})$. These two parameters essentially contain all the information
about the EOS of neutron star matter that can be extracted from GW observations.
It would be interesting to study the map between $(a_{\rm EOS},b_{\rm EOS})$ and various parametrizations
of the EOS introduced in the literature~\cite{Read:2008iy}.

\item
In this paper we have focused on BNS systems, but our formalism can be used for discussing
measurability of tidal parameters in mixed black hole-neutron star (BHNS) binary systems. 
However, while we have seen that for an equal-mass BNS systems the tidal dephasing at contact
was of order $-10$~rad (as analytically given by the approximate formula Eq.~\eqref{eq:psi_cont}),
for realistic mixed BHNS systems this dephasing turns out to be {\it smaller} by 
approximately two orders of magnitude.
More precisely, when considering BHNS systems with mass ratio
$q\equiv M_B/M_A>1$ the tidal dephasing at contact (formally defined by Eq.~\eqref{eq:x_contact})
can be written as the product of the equal-mass result~\eqref{eq:psi_cont} by a factor $F_A$,
\be
\Psi_{\rm BHNS_{Newt}}^{\rm contact}=-\dfrac{39}{32}\dfrac{k_2^A}{\C_A^{5/2}} F_A,
\ee
where the supplementary factor $F_A$ is given by
\be
F_A=\dfrac{4}{13}\dfrac{1+12 q}{q\sqrt{1+q}}\dfrac{1}{\left(1+q \,\C_A \C_B^{-1} \right)^{5/2}}
\ee
with $\C_B\equiv 1/2$ denoting the compactness of the black-hole and $\C_A\sim 1/6$ that of the NS. 
For instance, for a canonical BHNS system with $M_A=1.4M_\odot$, $\C_A=1/6$ and $M_B=10M_\odot$, 
so that $q\approx 7.14$, this formula predicts that the BHNS tidal dephasing at contact 
will be smaller than the BNS one by a factor $1/16$, so that the BNS, 10-radians tidal 
signal at contact becomes reduced to a level $\sim-0.6$ rad.
This reduction by more than one order of magnitude of the inspiral tidal dephasing signal
indicates that (even if the coherent analysis suggested above allows one to work with large
effective SNRs) it will probably not be possible for LIGO-Virgo observations to extract a useful 
measurement of $G\mu_2$ from the inspiral signal.
Note that the factor $F_A$ is a strongly decreasing function of $q$. 
For instance, for the extreme mass ratio $q=3$ (corresponding to a $4.2M_\odot+1.4M_\odot$ system),
the factor $F_A$ is equal to 0.33, leading to a maximum dephasing of order $-3.3$~rad. 
In addition, we should remember that the measurability of $G\mu_2$ effectively uses
only a rather small fraction of the total tidal dephasing signal. 
As discussed in Sec~\ref{sec:correlations}, this small fraction is
due to the fact that only a small part of $\Psi^T(f)$ is ``orthogonal'' to the signals
associated to the nontidal parameters $\lambda_a$. [For instance in the equal-mass BNS systems
discussed above only about 0.1~rad of the maximum 10~rad were useful in determining
the measurability of $G\mu_2$, see Eq.~\eqref{eq:psi_useful}].

\item Though our work has been using several simplifications (analytical approximations to both the 
nontidal and tidal part of the EOB Fourier domain phasing, Fisher matrix analysis), the various 
checks we have done make us confident that our main conclusions are robust. We leave to future work 
a fuller study using better approximations (direct Fourier transform of the full, numerically-generated 
EOB waveform, Monte-Carlo estimate of statistical errors,$\dots$).

\item Concerning our (binary-pulsar-data based) assumptions about the relative smallness of the spin
parameters $(\beta,\sigma)$, they will be verifiable (or falsifiable) once BNS inspiral signals with
sufficiently high SNRs become available: both through consistency checks of various measurements of 
tidal parameters, and through direct (Bayesian) analysis of the preferred a posteriori 
values of $(\beta,\sigma)$.

\end{itemize}

\acknowledgments
LV thanks IHES for hospitality during the development of this work.
We are grateful to N.~Chamel for giving us access to the tabulated
EOS BSK19, BSK20 and BSK21. The tables for EOSs GNH3, SLy, FPS and 
APR are from the LORENE C++ library (http://www.lorene.obspm.fr).
We thank F.~Pannarale for exchanging information about tabulated EOSs 
and S.~Bernuzzi for help in implementing the Hermite polynomials 
interpolation needed to deal accurately with tabulated EOS.

\appendix

\section{Description of tidal effects in the effective one body model}
\label{sec:eob}
This Appendix is devoted to collect all the technical elements that 
are needed to include the description of tidal effects in the EOB
formalism. The aim of presenting this (partly well known) material here, 
in a self-contained form, is twofold: first, it serves to give the reader the necessary
background to understand how the EOB curve of Fig.~\ref{fig:EOB_vs_PN_phase} 
was generated; second, it is a compact reminder of useful formulas in order 
to help the interested reader to do his own EOB implementation.

\subsection{Dynamics and waveforms}
\label{sec:eob_theory}
The EOB formalism~\cite{Buonanno:1998gg,Buonanno:2000ef,Damour:2001tu}
replaces the PN-expanded two-body interaction Lagrangian (or
Hamiltonian) by a resummed Hamiltonian, of a specific form, which
depends only on the relative position and momentum of the binary
system. For a nonspinning BBH system, it has been
shown that its dynamics, up to the 3PN level, can be described by the
following EOB Hamiltonian (in polar coordinates, within the plane of
the motion): 
\begin{equation}
\label{eq:Heob}
H_{\rm EOB}(r,p_{r_*},p_\varphi) \equiv M c^2\sqrt{1+2\nu (\hat{H}_{\rm eff}-1)} \  ,
\end{equation}
where
\begin{equation}
\label{eq:Heff}
\hat{H}_{\rm eff} \equiv \sqrt{p_{r_*}^2 + A(r) \left( 1 +
  \frac{p_{\varphi}^2}{r^2} 
+ z_3 \, \frac{p_{r_*}^4}{r^2} \right)} \, .
\end{equation}
Here $M\equiv M_A + M_B$ is the total mass, $\nu \equiv M_A \, M_B /
(M_A + M_B)^2$ is the symmetric mass ratio, and $z_3 \equiv 2\nu
(4-3\nu)$. In addition, we are using rescaled dimensionless
(effective) variables, namely $r \equiv r_{AB}c^2 / GM$ and
$p_{\varphi} \equiv P_{\varphi}c / (GM_A M_B)$, and $p_{r_*}$ is
canonically conjugated to a ``tortoise'' modification of
$r$~\cite{Damour:2009ic}.

A remarkable feature of the EOB formalism is that the complicated,
original 3PN Hamiltonian (which contains many corrections to the basic
Newtonian Hamiltonian $\frac{1}{2} \, {\bm p}^2 - 1/r$) can be
replaced by the simple structure (\ref{eq:Heob})-(\ref{eq:Heff}),
whose two crucial ingredients are: (i) a ``double square-root''
structure $H_{\rm EOB} \sim \sqrt{1+\sqrt{{\bm p}^2 + \cdots}}$ and
(ii) the ``condensation'' of most of the nonlinear relativistic
gravitational interactions in one function of the (EOB) radial
variable: the basic ``radial potential'' $A(r)$. The
structure of the function $A(r)$ is rather simple at 3PN,
being given by
\begin{equation}
\label{eq:3.3}
A^{\rm 3PN} (r) = 1-2u+2 \, \nu \, u^3 + a_4 \, \nu \, u^4 \, ,
\end{equation}
where $a_4 = 94/3 - (41/32)\pi^2$, and $u \equiv 1/r =
GM/(c^2r_{AB})$\label{u=1/r}. It was recently found that an excellent description
of the dynamics of BBH systems is obtained~\cite{Damour:2009kr} by:
(i) augmenting the presently computed terms in the PN
expansion (\ref{eq:3.3}) by additional 4PN and 5PN terms;
(i) Pad\'e-resumming the corresponding 5PN ``Taylor''
expansion of the $A$ function. In other words, the BBH (or ``point
mass'') dynamics is well described by a function of the form
\begin{equation}
\label{eq:3.4}
A^0(r) = P^1_5\left[1-2u+2\nu u^3 + a_4 \nu u^4 + a_5\nu u^5 + 
a_6\nu u^6\right]  ,
\end{equation}
where $P^n_m$ denotes an $(n,m)$ Pad\'e approximant. It was found in
Ref.~\cite{Damour:2009kr} (and then substantially confirmed by Ref.~\cite{Pan:2011gk}) 
that a good agreement between EOB and numerical-relativity 
BBH waveforms is obtained in an extended ``banana-like'' 
region in the $(a_5,a_6)$ plane approximately spanning
the interval between the points $(a_5,a_6)=(0,-20)$ and
$(a_5,a_6)=(-36,+520)$. In this work we will select the values
$a_5=-6.37$, $a_6=+50$, which lie within this region
(the use of other values within the ``good BBH fit''
region would have no measurable influence on the dynamics
in the presence of dynamical tidal effects).

The proposal of Ref.~\cite{Damour:2009wj} for including dynamical
tidal effects in the conservative part of the dynamics consists in
simply using Eqs.~\eqref{eq:Heob}-\eqref{eq:Heff} with the following
tidally-augmented radial potential 
\begin{equation}
\label{eq:A}
A(u) = A^0(u) + A^{\rm tidal} (u)  .
\end{equation}
Here $A^0(u)$ is the point-mass potential defined in
Eq.~\eqref{eq:3.4}, while $A^{\rm tidal}(u)$ is a supplementary
``tidal contribution'' of the form
\begin{align}
\label{eq:T9}
A^{\rm tidal}=\sum_{\ell\geq 2} - \kappa_\ell^T u^{2\ell+2}\hat{A}^{\rm tidal}_\ell(u)\,,
\end{align}
where the terms $\kappa^T_\ell u^{2\ell +2}$ represent the
leading-order (LO) tidal interaction, i.e., the Newtonian order tidal interaction. 

The additional factor $\hat{A}^{\rm tidal}_\ell(u)$ in Eq.~\eqref{eq:T9} 
represents the effect of distance dependent, higher-order relativistic
contributions to the dynamical tidal interactions: 1PN (first order in $u$,
or next-to-leading order), 2PN (of order $u^2$, or next-to-next-to-leading order) etc.
Here we will consider it written in the Taylor-expanded form
\begin{equation}
\label{eq:linear2PN}
\hat{A}^{\rm tidal}_\ell(u)=1+\bar{\alpha}_1^{(\ell)} u + \bar{\alpha}_2^{(\ell)} u^2 \ ,
\end{equation}
where $\bar{\alpha}_n^{(\ell)}$ are functions of $M_A$, $\C_A$,
and $k_\ell^A$ for a general binary and are  written as 
(see Eq.~(37) of~\cite{Damour:2009wj})
\be
\label{eq:bar_alpha1}
\bar{\alpha}_n^{(\ell)}\equiv \dfrac{\kappa_\ell^A\alpha^{A(\ell)}_n + \kappa_\ell^B\alpha^{B(\ell)}_n}{\kappa_\ell^A + \kappa_\ell^B}
\ee
where the $\alpha^{A(\ell)}_n$ is the coefficient of the $n$PN fractional 
correction to the tidal interaction potential of body $A$.
(see Sec.~IIIC of~\cite{Damour:2009wj}). 
The dimensionless coefficients $\alpha_n^{A(\ell)}$ is a function of
the dimensionless ratio $X_A\equiv M_A/M$.
The analytical expression of the ($\ell=2$) coefficient $\alpha_1^{A(2)}$
has been reported in~\cite{Damour:2009wj} 
(and then confirmed in~\cite{Vines:2010ca}) and reads
\be
\label{eq:alpha1}
\alpha_1^{A(2)}=\dfrac{5}{2}X_A,
\ee
which in the equal-mass case, $X_A=1/2$, yields $\bar{\alpha}^{(2)}_1=1.25$. 
Reference~\cite{Bini:2012gu} has succeeded in computing the first post Newtonian
octupolar ($\ell=3$) coefficient  $\alpha_1^{A(3)}$ and the {\it second post-Newtonian} 
quadrupolar ($\ell=2$) and octupolar ($\ell=3$) coefficients $\alpha_2^{A(\ell)}$.
We recall here only the most relevant, 2PN quadrupolar one, that reads
\be
\label{eq:alpha2}
\alpha_2^{A(2)}= \dfrac{337}{28}X_A^2 + \dfrac{1}{8} X_A + 3.
\ee
In the equal-mass case, $X_A=1/2$, the values of these coefficients are 
$\alpha_1^{A(2)}=\bar{\alpha}^{(2)}_1=5/4=1.25$ and $\alpha_2^{A(2)}=\bar{\alpha}_2^{(2)}=85/14\approx 6.071429$. 
In the main text we have considered {\it only} the tidal
(1PN and 2PN) quadrupolar contributions, i.e. we have consider only the 
$\ell=2$ value in Eqs.~\eqref{eq:At} and~\eqref{eq:Ahat}. 
In Sec.~\ref{sec:effect_of_ell} below we will investigate the (small)
effect of the higher-$\ell$ tidal corrections for an equal-mass
BNS. To do so, we adopt the simplifying assumption that the 
higher--multipolar tidal--amplification factors 
$\hat{A}^{\rm tidal}_\ell(u)$, for $\ell>2$, are taken to coincide with
the $\ell=2$ one. This means that the EOB model that we will use
here  contains only $\alpha_1^{A(2)}$ and $\alpha_2^{A(2)}$ higher order
tidal parameters that are taken to replace the various  
$\bar{\alpha}_n^{(\ell)}$, with $\ell=\{2,3,4,\dots\}$, entering 
Eq.~\eqref{eq:linear2PN}, i.e. $\hat{A}_\ell^{\rm tidal}\equiv \hat{A}_2^{\rm tidal}$
fpr $\ell>2$. Since the main effect is due to the leading-order, Newtonian
prefactor $\kappa_\ell^T u^{2\ell+2}$, this simplifying choice does not change
in a relevant manner the conclusions of Sec.~\ref{sec:effect_of_ell}.

Now that we have reminded the important elements needed to build the tidally
extended EOB Hamiltonian, let us move to discuss how the point-mass EOB 
waveform $h_{\lm}^0$ is augmented by tidal contributions. 
Similarly to the additive tidal modification~\eqref{eq:A} to the $A$ potential, 
we will consider an {\it additive} modification of the waveform~\cite{Baiotti:2011am}, 
having the structure
\begin{equation}
\label{eq:4.1}
h_{\lm} =  h_{\lm}^0  + h_{\ell m}^{\rm tidal}\,.
\end{equation}
The point-mass contribution is explicitly given by~\cite{Damour:2008gu}
\begin{equation}
\label{eq:4.0}
h^0_{\lm} = c_{\ell+\epsilon}(\nu) h_{\lm}^{\prime\,(N, \epsilon)} S^{(\epsilon)} \widehat{h}_{\ell m}^{\rm tail} \rho^\ell_{\ell m}\widehat{h}_{\ell m}^{\rm NQC}\,,
\end{equation}
where $\epsilon := \pi(\ell +m) = 0,1$ is the parity of the considered multipole,
where the coefficient
\be
c_{\ell+\epsilon}(\nu) = X_B^{\ell + \epsilon -1}+(-)^{\ell +\epsilon} X_A^{\ell +\epsilon -1},
\ee
has been separated off the Newtonian waveform $ h_{\lm}^{(N, \epsilon)} \equiv c_{\ell+\epsilon}(\nu) h_{\lm}^{\prime\,(N, \epsilon)}$, 
and where the other factors respectively represent: a source factor $S^{(\epsilon)}$, 
a tail factor $\widehat{h}_{\ell m}^{\rm tail}$, a resummed 
modulus correction $\rho^\ell_{\ell m}$ , and a next-to-quasi-circular correction, $\widehat{h}_{\ell m}^{\rm NQC}$. 
The latter correction contains 
two next-to-quasi-circular parameters $(a_1,a_2)$ as in Ref.~\cite{Damour:2009kr}. 
[Since we will be dealing with equal-mass binaries, we fix 
$a_1=  -0.0439$ and $a_2=1.3077$, according to the EOB/NR comparison (for a BBH equal-mass system) 
of Ref.~\cite{Damour:2009kr}]. The tail factor introduced here is given by 
$\widehat{h}_\lm^{\rm tail}=T_\lm e^{\ii\delta_\lm}$,
according to the notation of Ref.~\cite{Damour:2008gu}.
Using the recent computation~\cite{Vines:2011ud} of the
1PN-accurate Blanchet-Damour mass quadrupole moment~\cite{Blanchet:1989ki} 
of a tidally interacting binary system (together with the Newtonian-accurate spin 
quadrupole and mass octupole) and transforming their symmetric-trace-free 
tensorial results into our $\lm$-multipolar form, we have computed the corresponding
1PN-accurate value of $h_{22}^{\rm tidal}$, as well as the 0PN-accurate values of
$h_{21}^{\rm tidal}$, $h_{33}^{\rm tidal}$, and $h_{31}^{\rm tidal}$. 
In addition, using the general analysis of tail effects in 
Refs.~\cite{Blanchet:1992br,Blanchet:1995fr} and the 
resummation of tails introduced in 
Refs.~\cite{Damour:2007xr,Damour:2007yf,Damour:2008gu},
we were able to further improve the accuracy of these 
waveforms by incorporating (in a resummed manner) the effect 
of tails (to all orders in $M$). From a PN point of view, this 
means, in particular, that the  tidal contribution we use to the
total metric waveform is 1.5PN accurate. 
From the results of Ref.~\cite{Vines:2011ud}, 1PN source moment, one has that 
the only nonvanishing multipolar tidal corrections at 1PN fractional level are 
$h_{2m}^{\rm tidal}$ ($m=1,2$) and $h_{3m}^{\rm tidal}$ ($m=1,3$).
These multipolar components of the waveform can be obtained simply by computing the 
corresponding number of time derivatives (in the circular approximation)
of the multipole moments given by~\cite{Vines:2011ud} and then projecting 
them along symmetric-trace-free tensor spherical harmonics.
For consistency with the fact that tidal effects are included at 2PN fractional
accuracy in the Hamiltonian, we similarly write the (multipolar) waveform such
to formally include (yet analytically uknown) 2PN tidal corrections.
The $\ell=m=2$ tidal part of the waveform is written as 
\begin{align}
\label{eq:h22}
&h_{22}^{\rm tidal} = h_{22}^{\prime (N,0)} \nonumber\\
                  &\times\bigg\{\kappa_2^A\left(\dfrac{X_A}{X_B}+ 3 \right)v_\Omega^{10} \left[1+\beta_1^{22}(X_A) v_\Omega^2+\beta_2^{22}(X_A)v_\Omega^4\right] \nonumber\\
                  &+ \kappa_2^B\left(\dfrac{X_B}{X_A}+3\right)v_\Omega^{10}\left[1+\beta_1^{22}(X_B)v_\Omega^2+\beta_2^{22}(X_B)v_\Omega^4\right]\bigg\},
\end{align}
where the $\beta_1^{22}(X)$ and $\beta_2^{22}(X)$ functions parametrize respectively 
1PN and 2PN fractional tidal corrections. The 1PN tidal coefficient, $\beta_1^{22}(X)$, 
can be computed  analytically from the results of Ref.~\cite{Vines:2011ud} and 
it reads 
\be
\label{eq:beta1_22}
\beta_1^{22}(X) = \dfrac{-202 + 560 X - 340 X^2 + 45 X^3}{42(3 - 2 X)}.
\ee
The 2PN tidal coefficient $\beta_2^{22}(X)$ is currently unknown analytically.
Note that, following the original suggestion of Ref.~\cite{Damour:2006tr}, 
in Eq.~\eqref{eq:h22} we have replaced the PN ordering parameter $x=(M\Omega)^{2/3}$ 
by the EOB velocity variable $v_\Omega = r_\Omega \Omega$, where $r_\Omega=r\psi^{1/3}$
and $\psi$ is computed using the 3PN-accurate EOB Hamiltonian following the definition
(originally at 2PN accuracy) of Ref.~\cite{Damour:2006tr} (see also Ref.~\cite{Damour:2009ic}). 
The other waveform multipoles that present tidal corrections (up to the 2PN formal level) are
\begin{align}
h_{21} & = h^{\prime\,(N,1)}_{21}\bigg\{c_{2+1}(\nu) \hat{h}_{21} \nonumber\\
      &   + \left(\dfrac{9}{2}-6 X_B\right)v_\Omega^{10} \left[ 1+\beta_2^{21}(X_B)v_\Omega^2\right]\kappa^B_2 \nonumber\\
      &   - \left(\dfrac{9}{2}-6 X_A\right)v_\Omega^{10} \left[ 1+\beta_2^{21}(X_A)v_\Omega^2\right]\kappa^A_2  \bigg\}, \\
h_{3m} & = h_{3m}^{\prime\,(N,0)}\bigg\{c_{3+0}(\nu) \hat{h}_{3m} \nonumber\\
      &  + 6 X_A v_\Omega^{10}\left(1 + \beta_1^{3m}(X_B) v_\Omega^2\right)\kappa^B_2\nonumber\\ 
      &  - 6 X_B v_\Omega^{10}\left(1 + \beta_1^{3m}(X_A) v_\Omega^2\right) \kappa^A_2  \bigg\}.
\end{align}
where $\beta_1^{\ell m}(X)$ formally indicate the (currently unknown) corresponding 2PN corrections.
The tidally-corrected radiation reaction ${\cal F}_\varphi$ is then computed by using this corrected waveform 
in the definition of ${\cal F}_\varphi$ given in~\cite{Damour:2008gu,Damour:2009kr}. 
Note that the point-mass partial multipolar amplitudes $\rho_\lm$ that we use in the 
construction  of the analytic radiation reaction in this paper are augmented with respect 
to those discussed in Ref.~\cite{Damour:2008gu,Damour:2009kr} by the new (5PN accurate) 
$\nu=0$ terms recently computed in~\cite{Fujita:2010xj}. 
By contrast, for simplicity and for consistency with Ref.~\cite{Damour:2009kr},
we adopt the same prescription of that reference, Eq.~(4) there, to compute the $\ell=m=2$ 
point-mass waveform, which relies on a different resummation of the residual amplitude
correction with respect to Eq.~\eqref{eq:4.0} above. More precisely, the residual modulus 
correction of Eq.~\eqref{eq:4.0}, $(\rho_{22})^2$, is  replaced by the 
Pad\'e-resummed function $f_{22}^{\rm Pf}(x;\nu)=P^3_2[f_{22}^{\rm Taylor}(x;\nu)]$, 
where $f_{22}^{\rm Taylor}$ is computed in Ref.~\cite{Damour:2008gu} at $3^{+2}$ PN accuracy.
In addition, the residual phase correction $\delta_{22}$ that is used here is computed
at the accuracy given in~\cite{Damour:2008gu}, Eq.~(20), without the further ($\nu=0$) 
term obtained by~\cite{Fujita:2010xj}.

In summary, the EOB tidal model that we use here is {\it formally} complete up
to the 2.5PN level, though {\it analytically} complete at the 1.5PN level only, 
because of the (current) lack of analytical information on the coefficients 
$\{\beta_2^{22}(X),\,\beta_1^{21}(X),\,\beta_1^{31}(X),\,\beta_1^{33}(X)\}$.
Despite this, one should keep in mind that in the most relevant equal-mass case,
only the $\beta_2^{22}$ coefficient is relevant. On top of this, let us remind that 
in Sec.~\ref{sec:PNaccuracy} we argued that in the (Fourier domain) tidal phasing 
the contribution due to $\beta_2^{22}$ is very subdominant with respect to the others 
and thus it can be safely neglected in first approximation.
%
%
\begin{figure}[t]
\center
\includegraphics[width=0.5\textwidth]{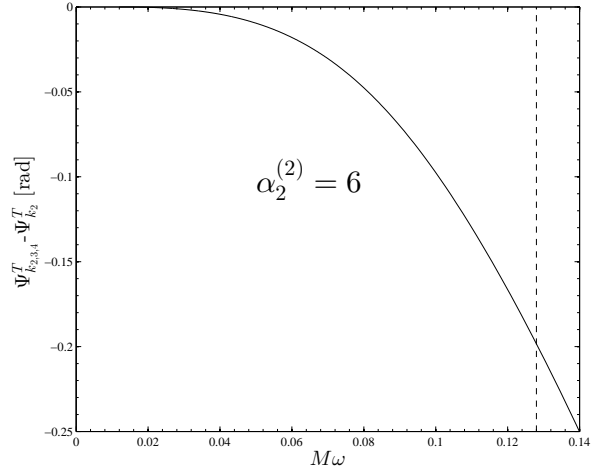}
\caption{\label{fig:l3l4}Difference between the full SPA tidal phase 
$\Psi^T_{k_{2,3,4}}$ obtained from a EOB waveform computed with
an $A^{\rm tidal}$ including $k_2$, $k_3$ and $k_4$, and 
the tidal phase computed with $k_2$ only. The figure refers to 
a $\gamma=2$, rest-mass polytrope BNS model with $\C=0.16$ 
and $\bar{\alpha}_2^{(2)}=6$. The vertical dashed line indicates 
the contact frequency.}
\end{figure}

In the main text we have used the EOB approach to compute a SPA-defined, Fourier
domain EOB tidal phasing, $\Psi^T_{\rm EOB_{SPA}}(\omega)$, where $\omega$ is the 
quadrupolar GW frequency  so as to {\it control} the accuracy of the Fourier-domain, 
PN-expanded tidal phase given by Eq.~\eqref{eq:2.5PN} for an equal-mass binary.
The analytical procedure to obtain  $\Psi^T_{\rm EOB_{SPA}}(\omega)$ (that we shall
henceforth simply denote as $\Psi^T(\omega)$) is described
in Sec.~\ref{EOB:SPA} (see in particular Eq.~\eqref{eq:EOBSPA}), while the 
comparison with the PN-expanded tidal phasing is discussed in Sec.~\eqref{sec:PNaccuracy}, 
in particular Figs.~\ref{fig:EOB_vs_PN_phase} and~\ref{fig:fig2} there.
The phase $\Psi^T(\omega)$ is computed by integrating numerically 
Eq.~\eqref{eq:PsiT_EOB} starting from the frequency $\omega_0$  that marks the 
beginning of the inspiral waveform obtained when solving the EOB equations of 
motion numerically. 
This integration is done using the 2.5PN result for $\Psi^T_{\rm 2.5PN}$  
and $d\Psi^T_{\rm 2.5PN}/d\omega$  as initial boundary conditions, and thus $\omega_0$ 
needs to be chosen  sufficiently small (i.e., the iniitla separation is 
sufficiently large) so to have $Q_\omega^{\rm 2.5PN}\approx Q_\omega^{\rm EOB}$.
In practice, for all compactnesses considered in Figs.~\ref{fig:EOB_vs_PN_phase} 
and~\ref{fig:fig2}, the initial relative separation that we use is $r_0=32$, 
which corresponds to a quadrupolar GW frequency $\omega_0\sim 0.111$.

\subsection{Dependence of the phasing on dynamical tidal effects with $\ell>2$ and its linearity with $\k$}
\label{sec:effect_of_ell}
%
%
\begin{figure}[t]
\center
\includegraphics[width=0.5\textwidth]{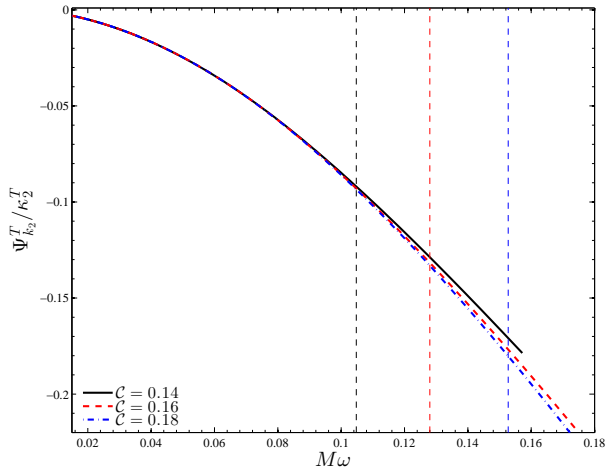}
\caption{\label{fig:linear_k2T}Approximate linearity of $\Psi^T_{\rm EOB}$ with respect 
to $\k$, with $\bar{\alpha}_2^{(2)}=6$. Three equal-mass (polytropic) BNS systems of 
compactnesses $\C=\{0.14,\,0.16,\,0.18\}$ are compared. The vertical dashed lines 
indicate the corresponding contact frequencies. }
\end{figure}
The use of the tidal phasing $\Psi^T_{\rm 2.5PN}$, Eq.~\eqref{eq:2.5PN} in the main text
neglects by construction two physical effects that are  incorporated in the 
EOB description, notably: (i) dynamical tidal terms with $\ell>2$ (e.g., $\ell=3$ and $\ell=4$)
that enter in the definition of $A^{\rm tidal}(u)$; (ii) nonlinear effects in $\kappa_2^T$,
that are present due to the resummed nature of the EOB formalism.
In this Section we analyze their influence of $\Psi^T(\omega)$ and argue
that they collectively contribute to the phasing by an amount of the order of $-0.2$~rad
up to contact. On top of this contribution being small, the fact that it is 
{\it negative} means that any measurability analysis neglecting it is on the conservative side.
Focusing on a $\C=0.16$, rest-mass, $\gamma=2$, polytrope binary, the effect 
of the  $\ell=3$ and $\ell=4$ tidal corrections to the EOB potential 
is illustrated in Fig.~\ref{fig:l3l4}.
The figure shows the difference between the full $\Psi^T_{k_{2,3,4}}$ obtained 
from a EOB waveform computed retaining $k_2$, $k_3$ and $k_4$ in the Hamiltonian
( through $A^{\rm tidal}$), and the phase $\Psi^T_{k_{2}}$ computed when retaining
only $k_2$.
One sees that the phase difference at the contact frequency 
(indicated by the dashed vertical line) is of the order of $-0.2$ rad.

Finally, to study to what extent the dependence of $\Psi^T_{k_2}$ is linear 
in $k_2$ (and therefore $\k$), we consider three binaries with compactnesses $\C=(0.14,0.16,0.18)$.
For each binary we compute the ratio $\Psi^T_{k_2}/\k$, that is displayed in Fig.~\ref{fig:linear_k2T}.
This figure indicates that $\Psi^T_{k_2}$ is linear in $\k$ to
good approximation. From the difference ($-0.0015$~rad) between the $\C=0.18$ 
and $\C=0.14$ curve at $M\omega=0.1048$ (contact of $\C=0.14$ binary) one can
estimate that taking into account the $\k$ would further decrease $\Psi^T_{k_2}$ by
no more than $-0.3$ rad. Again, neglecting this effects means that our estimates
will be on the conservative side.

\smallskip
\smallskip
\section{The tidal phase for a general binary at 2.5PN accuracy}
\label{sec:PN_general}
In this Appendix we collect PN-expanded expressions for
$Q_\omega^T$ and for $\Psi^T(\omega)$ for a general binary.
Such a result is here expressed as a function of the PN
ordering parameter $x$, of $X_A=M_A/M$ and $X_B=1-X_A$ and
of the dimensionless tidal parameter $\kappa_2^A$ defined
in Eq.~\eqref{eq:def_kappaA}. Note that this result is general,
in the sense that it holds for a neutron star binary of any
mass ratio, or a mixed, black-hole neutron star binary.
In this latter case the tidal parameter of one of 
the two objects is put to zero~\cite{Damour:2009vw}.
The equal-mass analytic expressions used in the main
text are obtained as a particular case of the equations
listed below, i.e., $X_A=X_B=1/2$. 

In detail, the 2.5PN accurate tidal part of the $Q_\omega(x)$ function, 
$Q_\omega^T(x)$, has the structure
\be
Q_\omega^T = Q_\omega^A(x) \kappa_2^A + Q_\omega^B(x)\kappa_2^B
\ee
where the $Q_\omega^A(x)$ contribution is written as the following PN series
\be
Q_\omega^A = q^A_{\rm Newt} x^{5/2}\left(1 + \hat{q}_1^A x + \hat{q}_2^A x^{3/2} + \hat{q}_3^A x^2 + \hat{q}_4^A x^{5/2}\right),
\ee
where the coefficients read
\begin{widetext}
\begin{align}
q_{\rm Newt}^A& = -\dfrac{5}{24\nu}\left(12 + \dfrac{X_A}{X_B}\right)=-\dfrac{5}{24\nu}\dfrac{12-11X_A}{1-X_A},\\
\hat{q}_1^A       & = \dfrac{3179-919X_A-2286X_A^2+260 X_A^3}{48(12-11X_A)},\\ 
\hat{q}_2^A & = -4\pi ,\\
\hat{q}_3^A & = \dfrac{1}{11X_A-12}
\bigg\{\dfrac{67702048 X_A^5-223216640 X_A^4+337457524 X_A^3-141992280 X_A^2+96008669 X_A-143740242}{338688}\nonumber\\
             &+(2X_A-3)\beta_2^{22}(X_A)
             +\left(\dfrac{13}{24}X - \dfrac{3}{4}X^2 + \dfrac{X^3}{3} -\dfrac{1}{8}\right)\beta_{1}^{21}(X_A)\nonumber\\
             &-(1-2X_A)(1-X_A)^2
             \left( \dfrac{1}{1344}\beta_{1}^{31}(X_A) + \dfrac{3645}{448}\beta_1^{33}(X_A)\right)
\bigg\},\\
\hat{q}_4^A & =  -\dfrac{\pi\left(27719-22127X_A+7022X_A^2-10232X_A^3\right)}{96(12-11X_A)},
\end{align}
\end{widetext}
and we left explicit the dependence on the (yet unknown) parameters entering the tidal contribution to
the waveform at fractional 2PN order, $\{\beta_2^{22},\,\beta_1^{21},\,\beta_1^{31},\,\beta_1^{33}\}$.
The tidal part of the phase is then written in the following form
\be
\Psi^T_{\rm 2.5PN} = \Psi_A^T(x) \kappa_2^A + \Psi_B^T(x)\kappa_2^B,
\ee
where the $\Psi_A^T(x)$ contribution is given by the following PN series
\be
\Psi_A^T = p^A_{\rm Newt} x^{5/2}\left(1 + \hat{p}_1^A x + \hat{p}_2^A x^{3/2} + \hat{p}_3^A x^2 + \hat{p}_4^A x^{5/2}\right),
\ee
whose coefficients read
\begin{widetext}
\begin{align}
p_{\rm Newt}^A & = -\dfrac{3}{16\nu}\left(12 + \dfrac{X_A}{X_B}\right)=-\dfrac{3}{16\nu}\dfrac{12-11 X_A}{1-X_A},\\
\hat{p}_1^A  & = \dfrac{5\left(3179 - 919X_A - 2286X_A^2 + 260 X_A^3\right)}{672(12-11X_A)},\\
\hat{p}_2^A  & = -\pi,\\
\hat{p}_3^A  & = \dfrac{1}{12-11X_A}\bigg\{\dfrac{39927845}{508032} 
                 - \dfrac{480043345}{9144576} X_A
                 + \dfrac{9860575}{127008}     X_A^2
                 - \dfrac{421821905}{2286144}  X_A^3
                 + \dfrac{4359700}{35721}      X_A^4
                 - \dfrac{10578445}{285768}    X_A^5\nonumber\\
             &   + \dfrac{5}{9}\left(1-\dfrac{2}{3}X_A\right)\beta_2^{22}
                 +\dfrac{5}{648}\left(3-13X_A+18 X_A^2 - 8X^3_A\right)\beta_1^{21}(X_A) \nonumber\\
             &   +\left(1-X_A\right)^2(1-2X_A)\left(\dfrac{5}{36288}\beta_1^{31}(X_A)+\dfrac{675}{448}\beta_1^{33}(X_A)\right)
             \bigg\},\nonumber\\
\hat{p}_4^A  & = -\dfrac{\pi(27719-22127X_A + 7022X_A^2 - 10232 X_A^3)}{672(12-11X_A)}.
\end{align}
\end{widetext}
The 1PN coefficient $\hat{p}_1^A$ above was obtained in Eqs.~(3.8a)-(3.8b) of~\cite{Vines:2011ud}.
Then the relation
\be
\kappa_2^A = 3 \dfrac{G\mu_2^A}{M^5}\dfrac{ X_B}{X_A},
\ee
(and similarly with $A\leftrightarrow B$) is used so to express the tidal phase in terms of the 
tidal polarizability coefficients $G\mu_2^A$ and $G\mu_2^B$. By defining
\be
G\bar{\mu}_2 
 \equiv \frac{1}{26} \left[\left(1 + 12 \frac{X_B}{X_A}\right)G \mu_{2}^A + \left(1 + 12 \frac{X_A}{X_B}\right) G\mu_{2}^B\right],
\ee
a straightforward calculation finally yields the tidal part of the phase in the form
\be
\Psi^T_{\rm 2.5PN}=-\dfrac{117G\bar{\mu}_2}{8\nu M^5}x^{5/2}\hat{\Psi}^T_{\rm 2.5PN}
\ee
that is the one used (with $\beta_2^{22}=\beta_1^{21}=\beta_1^{13}=\beta_1^{33}=0$) in Eq.~\eqref{eq:psi_T}
to perform the Fisher-matrix-based estimate of the measurability of $G\bar{\mu}_2$.

\bibliography{refs}

\end{document}